\documentclass[11pt,a4paper]{article}
\pdfoutput=1

\usepackage{bm}
\usepackage{amsmath,amssymb}
\usepackage{cite}

\usepackage{graphicx}
\usepackage{color}
\usepackage{upgreek}

\usepackage{hyperref} %%put this package declaration last to make work since it redefines things!!%%
\hypersetup{colorlinks=true, citecolor=blue, filecolor=black, linkcolor=blue, urlcolor=blue, pdfpagemode=UseNone}
\numberwithin{figure}{section}
\numberwithin{equation}{section}

\newcommand{\be}{\begin{equation}}
\newcommand{\ee}{\end{equation}}
\newcommand{\bea}{\begin{eqnarray}}
\newcommand{\eea}{\end{eqnarray}}
\def\beal#1\eeal{\begin{align}#1\end{align}}   %one & only for aligning
\def\besp#1\eesp{\begin{multline}#1\end{multline}} %split an equation with first line left aligned & later right aligned

\newcommand{\TRM}[1]{#1} %{{\color{blue}\bf #1}} %%comment out when finished!
\newcommand{\trm}[1]{#1}  %%changes made after arXiv but before CQG
\newcommand{\notes}[1]{}
\newcommand{\com}[1]{}

\newcommand{\Op}{\mathcal{O}}
\newcommand{\ph}{\varphi}
\newcommand{\vp}{\varphi}

\newcommand{\vpi}{\uppi}

\newcommand{\Lp}{a\Lambda_\p}

\newcommand{\p}{\mathrm{p}} %\newcommand{\p}{\sigma}
\newcommand{\dd}[2]{\delta_{\!\phantom{(} #1}^{\!(#2)}\!(\ph)}
\newcommand{\ddp}[3]{\delta_{\!\phantom{(} #1}^{\!(#2)}\!(#3)}

\newcommand{\ff}{\mathfrak{f}}

\newcommand{\htot}{H}

\newcommand{\prop}{\triangle}

\newcommand{\hs}{\hat{s}}
\newcommand{\q}{\text{q}}
\newcommand{\cl}{\text{cl}}

\newcommand\ie{\textit{i.e.}\ }
\newcommand\eg{\textit{e.g.}\ }
\newcommand\cf{\textit{cf.}\ }

\newcommand{\aka}{{a.k.a.}\ }
\newcommand{\etc}{{\it etc.}\ }
\newcommand{\viz}{{\it viz.}\ }
\newcommand{\half}{\tfrac{1}{2}}

\newcommand{\eps}{\varepsilon}

\newcommand{\morri}{Morris:2018mhd}
\newcommand{\morrii}{Morris:2018axr}
\newcommand{\yuji}{Igarashi:2019gkm}
\newcommand{\nn}{\nonumber}
\newcommand{\propH}{\prop_\Lambda}
\newcommand{\proph}[1]{\prop_{#1}}

\newcommand{\cu}[1]{\!#1\!}

\newcommand{\cG}{\check{\Gamma}}

\newcommand{\Po}{\mathcal{P}}

\begin{document}

\begin{titlepage}
%\begin{flushright}
%%{\tt hep-ph/yymmnn}
%{\tt SHEP xx-xx}
%\end{flushright}

\begin{center}
{\huge \bf The continuum limit of quantum gravity at second order in perturbation theory}

%Renormalizable quantum gravity and diffeomorphism invariance}

%of the conformal sector in quantum gravity}
%Relevant directions for the conformal factor in perturbative quantum gravity 
%\emph{or: Through the conformal factor to(wards) perturbatively renormalizable quantum gravity }} 
%\vskip.3cm
%{\huge \bf  and etc} 
\end{center}
\vskip1cm

%\title{xxx}
%\author{Tim R. Morris}

\begin{center}
{\bf Matthew Kellett, Alex Mitchell and Tim R. Morris}
\end{center}

%\affiliation{
\begin{center}
{\it STAG Research Centre \& Department of Physics and Astronomy,\\  University of Southampton,
Highfield, Southampton, SO17 1BJ, U.K.}\\
\vspace*{0.3cm}
{\tt  M.P.Kellett@soton.ac.uk, A.Mitchell-Lister@soton.ac.uk, T.R.Morris@soton.ac.uk}
\end{center}

\abstract{We show that perturbative quantum gravity based on the Einstein-Hilbert action, has a novel continuum limit. The renormalized trajectory emanates from the Gaussian fixed point along (marginally) relevant directions but enters the diffeomorphism invariant subspace only well below a dynamically generated scale. We show that for pure quantum gravity to second order in perturbation theory, and with vanishing cosmological constant, the result is the same as computed in the standard quantisation. Although this case is renormalizable at second order for kinematic reasons, the structure we uncover works in general. One possibility is that gravity has a genuine consistent continuum limit even though it has an infinite number couplings. However we also suggest a possible non-perturbative mechanism, based on the parabolic properties of these flow equations, which would fix all higher order couplings in terms of Newton's constant and the cosmological constant.}

\vskip4cm
%%comment out when finished!
%\begin{center}\textit{\today}\end{center} %%standard package gives date%%
%\begin{center}\textit{\DTMnow}\end{center} %%datetime2 package gives time also%%

\end{titlepage}

\tableofcontents

%\newpage

\section{Introduction}
\label{sec:intro}

\TRM{If one follows the by--now--standard procedures of perturbative quantum field theory, 
then one finds that (four dimensional) quantum gravity suffers from the problem that it is not perturbatively renormalizable \cite{tHooft:1974toh,Goroff:1985sz,Goroff:1985th,vandeVen:1991gw}. At an operational level, this is reflected in the fact that infinitely many coupling constants are required to absorb ultraviolet divergences, new couplings appearing at each loop order. However this effect in non-renormalizable theories is just a symptom of a more fundamental problem. The fundamental problem is formulating a \emph{continuum limit}: a limit in which the ultraviolet cutoff is sent to infinity in such a way as to leave behind an interacting quantum field theory that is finite at physical scales. It is this problem that we set out to solve in this paper.

Since the work of Wilson and others \cite{Wilson:1973,Wegner:1972my}, this problem has been understood at a much deeper level.  Wilson's great insight was to recognise that the continuum limit follows from a particular form of flow of effective actions as some intermediate cutoff scale $\Lambda$ is lowered, the so-called \emph{renormalized trajectory} \cite{Wilson:1973}. It is a flow out of an ultraviolet fixed point along  \emph{relevant} (or marginally relevant) directions. These directions are eigenoperator solutions of the flow equation linearised about the fixed point.
%Linearising the flow equation around the fixed point, these (marginally) relevant directions are eigenoperator solutions with non-negative renormalization group (RG) eigenvalues. 
In the perturbative case the fixed point is just the Gaussian one, corresponding to free fields, and (marginally) relevant eigenoperators have mass dimension $\le d$, where $d$ is the spacetime dimension (for us $d=4$). All eigenoperators with dimension greater than $d$ are \emph{irrelevant} and  correspond to directions that fall into the fixed point. As a result, their couplings do not survive as separate couplings in the continuum limit.

From this Wilsonian perspective it is obvious without calculation that four-dimensional quantum gravity has no perturbative continuum limit. When the Einstein-Hilbert term is expanded in small 
fluctuations, $H_{\mu\nu}$, the result is an infinite series of interactions
\be 
\label{irrelevant}
\sim H^n \partial H \partial H \qquad (n\ge1)\,,
\ee
 \emph{all} of which are irrelevant operators (of dimension $n+4$). Thus the only perturbative continuum limit is the Gaussian fixed point itself, $\sim(\partial H)^2$, a theory of free gravitons.}

\begin{figure}[ht]
\centering
\includegraphics[scale=0.3]{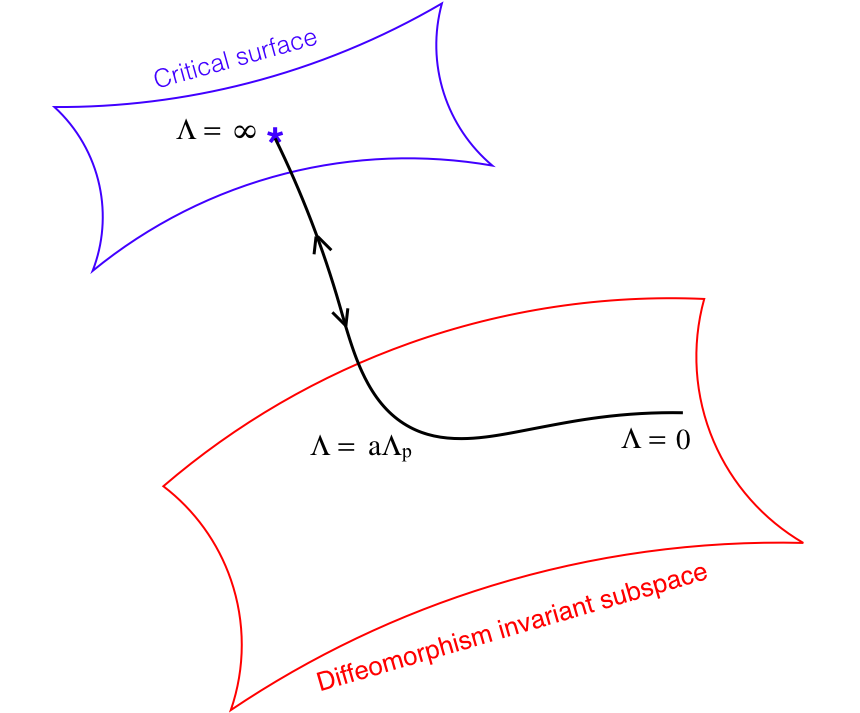}
%\vskip-30pt
\caption{The continuum limit is described by a renormalized trajectory that shoots out of the Gaussian fixed point (free gravitons) along (marginally) relevant directions that cannot respect diffeomorphism invariance for $\Lambda>a\Lambda_\p$, where $\Lambda_\p$ is a characteristic of the renormalized trajectory and is called the \emph{amplitude suppression scale}, and $a$ is a non-universal number. By appropriate choice of the \emph{underlying} couplings $g^\sigma_n$, diffeomorphism invariance is then recovered at scales $\Lambda,\vp\ll\Lambda_\p$ where also we recover an expansion in the \emph{effective} coupling  $\kappa\sim\sqrt{G}$.}
\label{fig:flow}
\end{figure}

\TRM{However in ref. \cite{\morri} we pointed out a subtle flaw in this reasoning. The conformal factor instability \cite{Gibbons:1978ac} renders meaningless Wilsonian RG \cite{Wilson:1973} flows involving otherwise arbitrary functions of the conformal factor amplitude, $\vp$, unless  they are expanded in the UV (ultraviolet) over a novel tower of increasingly relevant operators $\dd\Lambda{n}$ ($n=0,1,\cdots$) \cite{\morri}.  Exploiting this observation, in a series of papers we have been developing a continuum limit for quantum gravity that is perturbative in couplings but non-perturbative in Planck's constant, $\hbar$ \cite{\morri,Kellett:2018loq,Morris:2018upm,\morrii,first,secondconf}. 
%This new quantisation follows from the observation that, despite the conformal factor instability \cite{Gibbons:1978ac}, Wilsonian RG \cite{Wilson:1973} flows involving otherwise arbitrary functions of the conformal factor amplitude, $\vp$,  remain well defined but only if they are expanded in the UV (ultraviolet) over a novel tower of increasingly relevant operators $\dd\Lambda{n}$ ($n=0,1,\cdots$) \cite{\morri}. 
The result is the renormalized trajectory sketched in fig. \ref{fig:flow} \cite{first}. In this paper we complete the construction to second order, gaining further insight into the nature of this novel continuum limit.}

To the extent that other approaches address the continuum limit for quantum gravity, it has been tacitly assumed that this must take place within the \emph{diffeomorphism invariant subspace}, that is the space of actions that respect diffeomorphism invariance in some suitably well defined sense. 
%Here we mean diffeomorphism invariance expressed in some suitably well defined sense. 
For the Wilsonian RG, %this means that 
effective actions in this subspace must satisfy modified Slavnov-Taylor identities (mST) that encode the `breaking' of BRST invariance by the effective cutoff, $\Lambda$ \cite{Ellwanger:1994iz}. It turns out that with interactions built from $\dd\Lambda{n}$, it is only possible to respect these mST well below some characteristic scale, $\Lambda_\p$ \cite{\morrii}. Then the renormalized trajectory splits into two parts, a part above $a\Lambda_\p$ ($a$ is a non-universal number) that expresses the perturbative continuum limit by shooting out of the Gaussian fixed point along (marginally) relevant directions parametrised by an infinite number of (marginally) relevant underlying  couplings $g^\sigma_n$, and the part below $a\Lambda_\p$ where the renormalized trajectory enters the diffeomorphism invariant subspace. This part includes the physical amplitudes since they are recovered from the limit $\Lambda\cu\to0$. Here the trajectory becomes independent of the underlying couplings except indirectly through some collective effects, \aka diffeomorphism invariant effective couplings. In particular it is here that Newton's gravitational constant, $G$,  makes its appearance as expressed through %\TRM{$\kappa$}.
\be 
\label{kappa}
\kappa=\sqrt{32\pi G} \,.
\ee

\TRM{As we just stated, for cutoff scales $\Lambda\cu>\Lp$, where we construct the theory, there is no diffeomorphism invariance. Instead we work within some minimal general requirements. In particular, an essential ingredient for Wilsonian RG to make sense is the concept of Kadanoff blocking \cite{Kadanoff:1966wm}, where one integrates out degrees of freedom at short distances to obtain effective short range interactions. Thus we must work in Euclidean signature, so that short distance really does imply short range. Similarly for RG fixed points to exist, the manifold itself must look the same at any scale. That tells us to work with fluctuations on flat $\mathbb{R}^4$. Thus we construct the theory in flat Euclidean space.\footnote{\TRM{Although we stay with this case in this paper we note that, having constructed the theory in flat space, one can then study the construction on other manifolds and the analytic continuation to Lorentzian signature \cite{\morrii}.}} This means the metric is given by $\delta_{\mu\nu}$ and the interactions are constrained not by diffeomorphism invariance but only by Lorentz invariance (actually $SO(4)$ invariance). 

The Gaussian fixed point has free BRST invariance but the interactions we add do not respect its cohomology, that is until we enter the diffeomorphism invariant subspace. Once the cohomology is respected however, diffeomorphism invariance is then inevitable, as a consequence of the fact that (under broad assumptions) there is a unique deformation of the free BRST algebra  into an interacting one. This deformation is parametrised by the effective coupling $\kappa$.  At the classical level the uniqueness of this deformation is proven using the well developed subject of BRST cohomology \cite{Barnich:1993vg,Boulanger:2000rq} where one asks simultaneously for consistent deformations of linearised diffeomorphism invariance and local actions that realise it, whilst automatically taking into account all local field reparametrisations. This was generalised to the quantum level in ref. \cite{\morrii} and we continue to work within this formalism. It means for example, that the standard Riemannian measure $\sqrt{g}$ is not put in by hand  but appears as a consequence of the consistent solution of the BRST equations at higher order in $\kappa$, as we see explicitly in sec. \ref{sec:solCME}.
}

%This structure all logically follows from the conformal factor instability combined with the Wilsonian RG. The result however naturally resolves a number of conceptual puzzles. 
%%about applying RG concepts to quantum gravity. 
%In particular the notion of distance required in Kadanoff blocking, in the correlation length, and in conformal field theory (operator product short-distance expansion), is unclear now that the metric itself is quantised, see \eg \cite{Dietz:2015owa,Dietz:2016gzg,Ambjorn:2020rcn}. These issues disappear in this formulation since its quantisation necessarily happens ``off space-time'', as captured in the upper part of the renormalized trajectory. Distance in this upper part is defined by the fiducial Euclidean space which in fact must be flat $\mathbb{R}^4$ as required for the Wilsonian RG to make sense \cite{\morri}, and cannot be defined by the metric because it cannot be defined outside the diffeomorphism invariant subspace. In a similar way, the intuitive arguments against the existence of an ultraviolet fixed point in quantum gravity \cite{Aharony:1998tt,Shomer:2007vq}, are seen to be inapplicable since they are based on black hole entropy considerations \cite{first}.
In refs.  \cite{\morri,Kellett:2018loq,Morris:2018upm,\morrii,first} we established  this continuum limit to first order. In ref. \cite{secondconf} we established that an appropriate continuum limit exists at second order, in the sense that we showed that a well defined renormalized trajectory can be constructed, and that one can choose domains for the (marginally) relevant underlying couplings so that interactions satisfy certain \emph{trivialisation} conditions \cite{\morrii,first,secondconf} in the large $\Lambda_\p$ limit. These trivialisations  are necessary for the renormalized trajectory to enter the diffeomorphism invariant subspace  \cite{\morrii,first}. In this paper we complete the construction to second order, by verifying that the couplings can furthermore be chosen so that the mST is satisfied, and by computing the remainder of the renormalized trajectory. From this we also derive the physical Legendre effective action to $O(\kappa^2)$ by taking the limit $\Lambda\cu\to0$. We show that the result is identical to that derived in standard perturbative quantisation at one loop and second order in $\kappa$, where we choose to parametrise the metric in terms of fluctuations, $H_{\mu\nu}$, as 
%\TRM{in \eqref{gH}}.
\be 
\label{gH}
g_{\mu\nu} = \delta_{\mu\nu} +\kappa H_{\mu\nu}\,.
\ee

The second-order renormalized trajectory, being non-perturbative in $\hbar$,  involves a sum over tadpoles and melonic Feynman diagrams to all loops \cite{secondconf}. However on trivialisation in the large $\Lambda_\p$ limit, it collapses down to something that can be seen to be effectively one loop and second order in $\kappa$. 
We will see that undetermined parameters are left behind, associated to BRST invariant terms that run logarithmically with $\Lambda$, and furthermore these are the only places where such ambiguities appear. In this paper we only consider pure quantum gravity at vanishing cosmological constant. We show that  in this case  at this order, the logarithmic ambiguities are actually BRST exact, and can thus be absorbed in a wave-function-like canonical transformation. This is the BRST cohomological equivalent of the ``kinematical accident'' that pure gravity without cosmological constant, is one-loop finite in a  standard perturbative treatment \cite{tHooft:1974toh}. 

In sec. \ref{sec:discussion} we discuss the implications. It seems clear that even at $O(\kappa^2)$, once we add matter and/or a cosmological constant,  it will no longer be the case that logarithmic running inside the diffeomorphism invariant subspace is attributable to reparametrisation. One possibility then is that this construction ultimately leaves behind an infinite number of effective couplings that just correspond to the couplings that have to be added order by order in the number of loops in a standard perturbative treatment. The difference here however is that the result is a genuine continuum limit, \TRM{in the sense set out in the beginning paragraph}, apparently completely consistent, no matter how inconvenient this may seem phenomenologically. 

\trm{We emphasise that in this case the continuum limit \emph{exists}, but is controlled by an infinite number of (marginally) relevant, \emph{renormalizable}, underlying couplings. This situation should be contrasted with the situation in a non-renormalizable theory. In a non-renormalizable theory, it is impossible to construct a genuine continuum limit because the presence of fundamental (marginally) irrelevant couplings ensure that the critical surface cannot be reached in the limit that the cutoff $\Lambda$ is sent to infinity. Our result thus shifts the debate over quantum gravity. If the continuum limit of quantum gravity indeed has an infinite number of couplings, then the question of whether one regards that as acceptable is now a matter of taste or philosophy, rather than internal consistency of the theory itself, until such a time that such a theory of quantum gravity can be fully confronted with experiment.}

However in sec. \ref{sec:missing} we point to a possible novel non-perturbative mechanism that would fix all effective couplings in terms of just  Newton's constant and the cosmological constant. This is based on the 
%severely limited capacity for non-singular solutions that follows from the 
parabolic property of these flow equations. In the $\vp$ sector, flows are guaranteed well-defined only in the UV direction (backward parabolic property). This has already played a key r\^ole in the construction \cite{\morri,Kellett:2018loq,Morris:2018upm,\morrii,first,secondconf}. However once diffeomorphism invariance is imposed by the mST, the solution is necessarily non-polynomial in the graviton. Then it matters that in the graviton sector, the flow is forward parabolic, and thus is guaranteed well defined only in the IR (infrared) direction. Since in reality these sectors cannot be treated separately, we see that non-perturbatively we are dealing with novel partial differential equations whose solutions typically fail in whichever direction they are evolved. We argue, using a simple linearised  model, that if we ensure a solution in which $\kappa$ is freely variable, then only the cosmological constant can also be freely variable, because perturbations in the higher derivative couplings would lead to singular trajectories. 

\TRM{For the sake of simplicity, all the results in the previous papers \cite{\morri,Kellett:2018loq,Morris:2018upm,\morrii,first,secondconf} were derived in Feynman - De Donder gauge. For  the most part in this paper we also use this gauge. However in sec. \ref{sec:gauge}  we discuss how the picture changes in more general gauges.  In particular we emphasise that on-shell amplitudes, and correlators of BRST invariant operators more generally, are guaranteed to be independent of gauge parameter provided the renormalized trajectory enters the diffeomorphism invariant subspace. Therefore the only real question is how the trajectory for $\Lambda\cu>\Lp$
changes. We investigate the latter by working in general De Donder gauge. We show that the trajectory has the same properties for an infinite range of the gauge fixing parameter, and we discuss briefly the interpretation of new effects that open up in some finite range.}

The structure of the paper is as follows. In sec. \ref{sec:review}, we collect together the material we need from the previous papers in order to derive the solution for the renormalized trajectory inside the diffeomorphism invariant subspace. The general form of the second order renormalized trajectory is given in terms of a particular integral and a complementary solution \cite{secondconf}. The remaining freedom in the underlying couplings is held in the complementary solution. We need to choose it so that the second-order mST is satisfied. As we will see in sec. \ref{sec:evaluation}, this involves at intermediate stages solving the Zinn-Justin identities \cite{ZinnJustin:2002ru} for a local effective action and BRST charge, to one loop and $O(\kappa^2)$. This step can be treated separately and is done in sec. \ref{sec:solCME}. A key r\^ole will be played by BRST invariant bilinear terms up to fourth order in space-time derivatives. In sec. \ref{sec:BRSTexact} we establish that all such terms are actually BRST-exact and thus can be eliminated by a canonical transformation. We finish this section by uncovering some higher-derivative symmetries that appear at this level, whose significance is unclear to us. The main part of the paper is contained in sec. \ref{sec:evaluation} where we demonstrate that inside the diffeomorphism invariant subspace, the renormalized trajectory at second order collapses to a solution that is equivalent to one derived using a standard perturbative approach to one loop and $O(\kappa^2)$, albeit in terms of an effective action for quantum fields and regularising using an effective cutoff $\Lambda$. We solve both for the derivative expansion at finite $\Lambda$ (with the help of app. \ref{app:derivexp}) and for the physical vertices in the limit $\Lambda\cu\to0$. In sec. \ref{sec:discussion} we discuss the significance of the logarithmic running in the current context, in particular for the generalisations to higher order or/and when a cosmological constant or matter is added. In sec. \ref{sec:missing} we uncover hints of a non-perturbative mechanism that fixes higher order couplings, as already mentioned above. \TRM{Sec. \ref{sec:gauge} addresses gauge parameter independence as already discussed above.}
Finally, in sec. \ref{sec:conclusions} we summarise, make some further comments, and draw our conclusions.

\section{Review}
\label{sec:review}

In this paper we will solve for the physical Legendre effective action
\be 
\label{physical}
\Gamma_\text{phys} = \lim_{\Lambda\to0} \Gamma\,,
\ee
which corresponds to sending the IR (infrared) cutoff $\Lambda\cu\to0$. We do this by solving for the final part of the renormalized trajectory, $\Lambda\cu\ll a\Lambda_\p$, inside the diffeomorphism invariant subspace, \cf fig. \ref{fig:flow}. In this section we provide a brief review of the earlier research \cite{\morri,Kellett:2018loq,Morris:2018upm,\morrii,\yuji,first,secondconf}. This will also serve to collect together equations in a form that we will need later.
%the background material, including results needed later, and then further adapt it for this region. 
%$\Lambda\cu\ll a\Lambda_\p$.
The (IR cutoff) effective action 
%$\Gamma$ is expressed in terms of a free part, $\Gamma_0$, which includes the free BRST transformations, plus the interaction part $\Gamma_I[\Phi,\Phi^*]$:
\be 
\label{Gamma}
\Gamma = \Gamma_0 + \Gamma_I\,,
\ee
is expressed in terms of an interaction part $\Gamma_I[\Phi,\Phi^*]$
and a free part $\Gamma_0[\Phi,\Phi^*]$. \trm{Here $\Phi$ and $\Phi^*$ are the collective notation for the classical fields and antifields respectively. The classical fields ($H_{\mu\nu}, c_\mu$, and after gauge fixing, $\bar{c}_\mu$) are the expectation values in presence of sources that are the usual arguments of the Legendre effective action while, following Batalin and Vilkovisky  \cite{Batalin:1981jr,Batalin:1984jr}, antifields ($H^*_{\mu\nu}$, $c^*_\mu$) are introduced as sources for BRST transformations.} In the following, \trm{\emph{all expressions for actions should be understood as integrated over four flat Euclidean space-time dimensions}}. Thus,
in the minimal gauge invariant basis in which we work \cite{\yuji,first}, we write $\Gamma_0$ as:
%the free action is $\Gamma_0 = \cG^0_0+\cG^1_0$, 
%the free part is 
\be 
\label{Gzero}
\Gamma_0 = \half \left(\partial_\lambda \htot_{\mu\nu}\right)^2 -2 \left(\partial_\lambda \ph\right)^2 - \left(\partial^\mu \htot_{\mu\nu}\right)^2 +2\,\partial^\alpha\! \ph\, \partial^\beta \htot_{\alpha\beta} %\,,\qquad \cG^1_0 =  
\,- 2\,\partial_\mu c_\nu H^*_{\mu\nu}\,.
\ee
It is the action for free graviton fields $H_{\mu\nu}$ ($\vp=\half H_{\mu\mu}$) plus the antifield $H^*_{\mu\nu}$ source term for
\be 
\label{QH}
Q_0 H_{\mu\nu} = \partial_\mu c_\nu+\partial_\nu c_\mu\,,
\ee
the only non-vanishing free linearised BRST transformation in this basis,  %\cf \eqref{Gamma}, 
$c_\mu$ being the ghost fields. We introduce the \trm{Batalin-Vilkovisky} antibracket \cite{Batalin:1981jr,Batalin:1984jr,\yuji,first}, such that for functionals $\Xi[\Phi,\Phi^*]$ and $\Upsilon[\Phi,\Phi^*]$,
\be 
\label{QMEbitsPhi}
(\Xi,\Upsilon) = \frac{\partial_r\Xi}{\partial\Phi^A}\,\frac{\partial_l\Upsilon}{\partial\Phi^*_A}-\frac{\partial_r\Xi}{\partial\Phi^*_A}\,\frac{\partial_l\Upsilon}{\partial\Phi^A}\,.
%\qquad\mathrm{and}\qquad \Delta\, \Xi = (-)^{A+1} \frac{\partial_r}{\partial\Phi^A}\,\frac{\partial_r}{\partial\Phi^*_A}\,\Xi\,.
\ee 
\trm{where, again following Batalin and Vilkovisky, we are using the compact DeWitt notation.}
In $\Gamma_0$ we have chosen left-acting BRST transformations \cite{\morrii,first} so that the free BRST transformation is given by the first of the following equations:
\be 
\label{charges}
Q_0\, \Phi^A := (\Gamma_0,\Phi^A)\,,\qquad Q^-_0\Phi^*_A := (\Gamma_0,\Phi^*_A)\,.
\ee
Here we have taken the opportunity also to define the free Koszul-Tate operator $Q^-_0$. The superscript is a reminder that it lowers antighost number by one. From the definition \eqref{charges} and the free action \eqref{Gzero}, the non-vanishing free Kozsul-Tate differentials are:
\be 
\label{KTHc}
Q^-_0 H^*_{\mu\nu} = -2G^{(1)}_{\mu\nu}\,,\qquad Q^-_0 c^*_\nu = -2 \partial_\mu H^*_{\mu\nu}\,,
\ee
where $G^{(1)}_{\mu\nu}$ is the linearised Einstein tensor:
\be 
\label{Gmunu}
G^{(1)}_{\mu\nu} = -R^{(1)}_{\mu\nu}+\half R^{(1)}\delta_{\mu\nu} = \half\, \Box\, H_{\mu\nu} -\delta_{\mu\nu}\Box\,\vp+\partial^2_{\mu\nu}\vp+\half \delta_{\mu\nu}\partial^2_{\alpha\beta} H_{\alpha\beta}-\partial_{(\mu}\partial^\alpha H_{\nu) \alpha}\,,
\ee
the linearised curvatures being\footnote{defining symmetrisation as: $t_{(\mu\nu)} = \half (t_{\mu\nu}+t_{\nu\mu})$, and antisymmetrisation as $t_{[\mu\nu]} = \half (t_{\mu\nu}-t_{\nu\mu})$.}
\be 
\label{curvatures}
R^{(1)}_{\mu\alpha\nu\beta} = -2 \partial_{[\mu|\,}\partial_{[\nu} H_{\beta]\,|\alpha]}\,,\  R^{(1)}_{\mu\nu} = -\partial^2_{\mu\nu}\vp+\partial_{(\mu}\partial^\alpha H_{\nu) \alpha}-\half\, \Box\, H_{\mu\nu}\,,\  R^{(1)} = \partial^2_{\alpha\beta}H_{\alpha\beta}-2\,\Box\,\vp\,.
\ee
We also introduce the Batalin-Vilkovisky measure operator $\Delta$ \cite{Batalin:1981jr,Batalin:1984jr}, however it is regulated by a UV (ultraviolet) cutoff function $C^\Lambda(p)\equiv C(p^2/\Lambda^2)$ which satisfies $C(0)=1$ \cite{\morrii,\yuji,first}. Under anti-ghost grading, it splits into two parts that lower antighost number by one or two respectively ($\Delta^-$ simplifies to this in minimal basis \cite{\morrii}):
\be 
\label{measure}
\Delta = \Delta^- + \Delta^=\,,\qquad\Delta^- = \frac{\partial}{\partial H_{\mu\nu}}C^\Lambda\frac{\partial_l}{\partial H^*_{\mu\nu}}  \,,\qquad \Delta^= = - \frac{\partial_l}{\partial c_\mu}C^\Lambda\frac{\partial}{\partial c^*_\mu}\,.
\ee
The flow equation for $\Gamma_I$ takes the form \cite{Nicoll1977, Wetterich:1992, Morris:1993} (see also  \cite{Weinberg:1976xy,Morris:2015oca,Bonini:1992vh,Ellwanger1994a,Morgan1991}):\footnote{\trm{Notice the difference between  $\prop$ (triangle) and $\Delta$ (Delta):  the first is a propagator, and the second the measure operator.}}
\be %[18.9,10] = [6:8] is derived in p851 QG notes
\label{flow}
\dot{\Gamma}_I = -\half\, \text{Str}\left( \dot{\prop}_\Lambda\propH^{-1} \left[1+\propH \Gamma^{(2)}_I \right]^{-1}\right) 
%= -\half (-)^A \left( \dot{\prop}_\Lambda\propH^{-1}\left[1+\propH \Gamma^{(2)}_I \right]^{-1}\right)^{\!A}_{\ \ A} 
\,,
\ee
where the over-dot is $\partial_t =-\Lambda \partial_\Lambda$. The BRST invariance is  expressed through the mST  \cite{Ellwanger:1994iz,\yuji}:
\be 
\label{mST}
\half (\Gamma,\Gamma) - \text{Tr}\left( \!C^\Lambda\,  \Gamma^{(2)}_{I*} \left[1+\propH\Gamma^{(2)}_I\right]^{-1}\right) = 0\,.
\ee
In these equations we have introduced Str$\,\mathcal{M} = (-)^A\, \mathcal{M}^{A}_{\ \,A}$ and Tr$\,\mathcal{M} = \mathcal{M}^{A}_{\ \,A}$, and set
\be 
\label{Hessian}
\Gamma^{(2)}_{I\ AB} = \frac{\partial_l}{\partial\Phi^A}\frac{\partial_r}{\partial\Phi^B}\Gamma_I\,,\qquad \
\left(\Gamma^{(2)}_{I*}\right)^{A}_{\ \ B} 
\,=\,  \frac{\partial_l}{\partial\Phi^*_A}\frac{\partial_r}{\partial\Phi^B}\Gamma_I\,.
\ee 
The above equations are compatible \cite{Ellwanger:1994iz,\yuji} and both UV and IR (infrared) finite, the latter thanks also to the presence of the associated IR cutoff
%\footnote{These cutoffs were written as $C^\Lambda\equiv C$ and $C_\Lambda \equiv \bar{C}$ in refs. \cite{\morri,\morrii}.} 
$C_\Lambda= 1-C^\Lambda$, which appears in the IR regulated propagators as 
\be
\nonumber
\trm{
\propH^{AB} = C_\Lambda\prop^{AB}\,.
}
\ee
The cutoff function is chosen so that $C(p^2/\Lambda^2)\cu\to0$ sufficiently fast as $p^2/\Lambda^2\cu\to\infty$ to ensure that all momentum integrals are indeed UV regulated (faster than power fall off is necessary and sufficient). It is also required to be smooth (differentiable to all orders), corresponding to a local Kadanoff blocking. It thus permits for $\Lambda\cu>0$,
a solution for $\Gamma_I$ that has a space-time derivative expansion to all orders.
% quasi-local solution for $\Gamma_I$, namely one that has a space-time derivative expansion to all orders, as required 
We insist on this since it is equivalent to imposing locality on a bare action. Finally,
the propagators are defined (in Feynman - De Donder gauge) as follows \cite{\morrii}:
\beal 
\label{defs}
\prop^{AB}  &= \langle\Phi^A\,\Phi^B\rangle\,,\qquad
%\quad\text{and}\quad
\Phi^A(x) = \int_p  \text{e}^{-i p\cdot x}\, \Phi^A(p)\,,\qquad
\int_p \equiv \int\!\! %\frac{d^4p}{(2\pi)^4}\,.\\  
\frac{d^dp}{(2\pi)^d}\,.\\
\label{HH}
\langle H_{\mu\nu}(p)\,H_{\alpha\beta}(-p)\rangle &= \frac{\delta_{\mu(\alpha}\delta_{\beta)\nu}}{p^2}
-\frac1{d-2}
%\frac12
\frac{\delta_{\mu\nu}\delta_{\alpha\beta}}{p^2}\,, \\
\label{hh}
\langle h_{\mu\nu}(p)\,h_{\alpha\beta}(-p)\rangle &= \frac{\delta_{\mu(\alpha}\delta_{\beta)\nu}-
%\frac14\delta_{\mu\nu}\delta_{\alpha\beta}}{p^2} 
\frac1d\delta_{\mu\nu}\delta_{\alpha\beta}}{p^2} 
\,,\\
\label{pp}
\langle \ph(p)\,\ph(-p)\rangle &=  - \frac{d}{2(d-2)} 
\frac1{p^2}\,.\\
\label{cc}
\langle c_\mu(p)\, \bar{c}_\nu(-p)\rangle &= -\langle \bar{c}_\mu(p)\, c_\nu(-p) \rangle =  \delta_{\mu\nu}/{p^2}\,.
\eeal
Here $h_{\mu\nu}$ is the traceless part:
\be 
\label{h}
H_{\mu\nu} = h_{\mu\nu} + %\tfrac12\, \ph\, \delta_{\mu\nu}\,. 
\tfrac2d\, \ph\, \delta_{\mu\nu}\,.
\ee
All the above formulae apply to $d$ spacetime dimensions. 
%although $d$ does not explicitly enter in \eqref{cc}. 
For the most part we will work in the physical $d\cu=4$ dimensions, but later we will find it useful to employ dimensional regularisation as an intermediate step. 
Ghost propagator corrections are computed after shifting to gauge fixed basis using (in $d$ dimensions)
\be
\label{gaugeFixed}
%\bar{c}^*_\mu \,|_\text{gf}  &= \bar{c}^*_\mu \,|_\text{gi} + F_\mu\,,\\ \nonumber
H^*_{\mu\nu} \,|_\text{gi} = H^*_{\mu\nu} \,|_\text{gf} +\partial_{(\mu} \bar{c}_{\nu)} -\half\,\delta_{\mu\nu}\, \partial\!\cdot\! \bar{c}\,,
%\partial_\alpha\bar{c}_\alpha\,.
\ee
after which we shift back to gauge invariant basis \cite{\yuji,first}.

In the new quantisation, one expands $\Gamma_I$ perturbatively in its interactions, 
\be 
\label{expansion}
\Gamma_I = \sum_{n=1}^\infty\Gamma_n\,{\epsilon^n}/{n!} \,,
\ee
where $\epsilon$ is a formal small parameter, the true small parameter being the underlying couplings $g^\sigma_n$. However the treatment is non-perturbative in $\hbar$. 
At first order \eqref{flow} and \eqref{mST} become
\beal 
\label{flowone}
\dot{\Gamma}_1 &=  \half\, \text{Str}\, \dot{\prop}_\Lambda \Gamma^{(2)}_1 \,, \\
\label{mSTone}
0 &=  (\Gamma_0,\Gamma_1) - \text{Tr}\left( \!C^\Lambda\,  \Gamma^{(2)}_{1*} \right) = (Q_0+Q^-_0-\Delta) \Gamma_1 =: \hs_0\, \Gamma_1\,,
\eeal
\ie these equations are the linearised versions of the flow equation and mST. They play a fundamental r\^ole also at higher orders, since they govern the freedom in the solution at each order, \ie the form of the new interactions and their parameterisation in terms of couplings.

\subsection{Solutions to the linearised equations}
\label{sec:sollinear}

The first equation, \eqref{flowone}, is the flow equation satisfied by eigenoperators. As a result of the conformal factor instability, the eigenoperators we expand in are (integer $l\cu\ge0$ and $\eps=0(1)$ according the even(odd)  $\vp$-amplitude parity) \cite{\morri,\morrii,first}
\be 
\label{topFull}
\dd\Lambda{2l+\eps}\, \sigma(\partial,\partial\vp,h,c,\Phi^*) +\cdots\,,
\ee
so that  there is convergence of the sum over eigenoperators in the square-integrable sense \cite{\morri}. Here we have displayed the `top term',  $\sigma$ being a $\Lambda$-independent Lorentz invariant monomial involving some or all of the components indicated, in particular the arguments $\partial\vp,h,c,\Phi^*$  can appear as they are, or differentiated any number of times. The operators
\be
\label{physical-dnL}
\dd{\Lambda}{n} := \frac{\partial^n}{\partial\vp^n}\, \dd{\Lambda}{0}\,, \qquad{\rm where}\qquad \dd{\Lambda}0 := \frac{1}{\sqrt{2\pi\Omega_\Lambda}}\,\exp\left(-\frac{\vp^2}{2\Omega_\Lambda}\right)
\ee
have dimension $-1\cu-n$, and are responsible for turning gravity into a genuine perturbatively renormalizable quantum field theory. In these equations we have introduced
\be 
\label{Omega}
\Omega_\Lambda= |\langle \ph(x) \ph(x) \rangle | =  \int_q \frac{{C}(q^2/\Lambda^2)}{q^2} 
= \frac{\Lambda^2}{(4\pi)^2}\int^\infty_0\!\!\!\!\!\!du\, C(u) =
\frac{\Lambda^2}{2a^2}
\ee
which is the modulus of the $\ph$-tadpole integral (and $a$ a dimensionless non-universal number).
Since $\Omega_\Lambda$ is $O(\hbar)$ the operators are non-perturbative in $\hbar$ and this is the reason that the equations need to be treated non-perturbatively in $\hbar$. However expansion over these operators is the correct thing to do only in the UV regime, since the expansion converges if and only if $\Lambda\cu>a\Lambda_\p$. (This is actually the definition of $\Lambda_\p$ \cite{\morri,first}.) As we will see, in the IR regime we recover a sense in which the solutions can be expanded perturbatively in $\hbar$. Notice that undifferentiated $\vp$ does not appear in $\sigma$ but only in $\dd\Lambda{2l+\eps}$. The tadpole operator on the RHS of linearised flow equation  \eqref{flowone} generates a finite number of $\Lambda$-dependent UV regulated tadpole corrections involving fewer fields in $\sigma$ (and which vanish in the limit $\Lambda\cu\to0$). These are the terms we indicate with the ellipses. 

The general solution of the linearised flow equation \eqref{flowone} can be written as $\Gamma_1 \cu=\Gamma(\mu)$ where,
\be 
\label{complementary}
\Gamma(\mu) = 
\exp\left(-\frac12 {\prop}^{\Lambda\,AB} \frac{\partial^2_l}{\partial\Phi^B\partial\Phi^A}\right)\,
\Gamma_{\text{phys}}(\mu) \ =\ \sum_\sigma \left( \sigma f^\sigma_\Lambda(\vp,\mu)+\cdots \right)\,.
\ee
It is a linear sum over the eigenoperators \eqref{topFull} with constant coefficients, these being the  underlying couplings $g^{\sigma}_{2l+\eps}\!(\mu)$, and where this sum is subsumed into coefficient functions:
\be 
\label{coeffgen}
f^{\sigma}_\Lambda(\vp,\mu) = \int^\infty_{-\infty}\frac{d\vpi}{2\pi}\, \ff^{\sigma}(\vpi,\mu)\, {\rm e}^{-\frac{\vpi^2}{2}\Omega_\Lambda+i\vpi\vp} \,, 
\qquad
\ff^{\sigma}(\vpi,\mu) = i^{\,\eps}\sum_{l=0}^\infty (-)^l g^{\sigma}_{2l+\eps}\!(\mu)\, \vpi^{\,2n+\eps}\,.
\ee
The tadpole corrections are those contributions formed by attaching propagators in \eqref{complementary} to $\sigma$ (either exclusively or also to $\vp$). It can be shown that  the Taylor series of $\ff^{\sigma}(\vpi,\mu)$ converges absolutely for all $\vpi$, and furthermore $\ff^{\sigma}(\vpi,\mu)$ decays exponentially for $\vpi > 1/\Lambda_\p$ \cite{\morri,first}. This solution therefore makes sense for all $\Lambda\cu\ge0$.

There are thus an infinite tower of underlying couplings associated to every monomial $\sigma$. 
At first order, the couplings can be regarded as $\mu$ independent, and it turns out that all are relevant except for one marginal coupling \cite{\morrii,first}. At higher order, new higher dimension monomials $\sigma$ appear through the quantum corrections. Infinitely many of their underlying couplings are also relevant, however  the first few are irrelevant. These latter are not freely variable but determined by the requirement that we have a well-defined renormalized trajectory \cite{\morri,secondconf}. At second order there are no new marginal couplings, the first order couplings still do not run, while the irrelevant couplings that now appear, are determined in terms of the first order couplings \cite{secondconf}. 

Despite this, both at first order and second order, the relevant couplings
%to be chosen so as to satisfy \eqref{mSTtwo} in the large amplitude suppression scale limit.
can be chosen so that the amplitude suppression scale of each $f^\sigma_\Lambda(\vp,\mu)$ is at a common value $\Lambda_\p$, independent of $\sigma$,  and such that these coefficient functions trivialise in a way that we will require in order to have a chance of satisfying the mST \eqref{mST} \cite{\morrii}. By trivialise we mean that they have limiting behaviour \cite{first}
\be 
\label{flat}
f^\sigma_\Lambda(\vp,\mu) \to A_\sigma\qquad\text{as}\quad\Lambda_\p\to\infty\,,
\ee
or more generally ($\alpha$ a non-negative integer),
\be 
\label{flatp}
f^{\sigma}_\Lambda(\vp,\mu) \to A_{\sigma} \left({\Lambda}/{2ia}\right)^\alpha H_\alpha\!\left({ai\vp}/{\Lambda}\right)
\qquad\text{as}\quad\Lambda_{\p}\to\infty\,,
\ee
or indeed vanish in this limit. Here $A_\sigma$ is a constant, and $H_\alpha$ is the $\alpha^\text{th}$ Hermite polynomial:
\be 
\label{Hermite}
\left({\Lambda}/{2ia}\right)^\alpha H_\alpha\!\left({ai\vp}/{\Lambda}\right) = \vp^\alpha + \alpha(\alpha-1)\,\Omega_\Lambda\vp^{\alpha-2}/2+\cdots\,.
\ee
This is the unique form for $f^\sigma_\Lambda(\vp,\mu)$ such that it satisfies the linearised flow equation \eqref{flowone} and becomes $\vp^\alpha$ in the physical ($\Lambda\cu\to0$) limit, the tadpole corrections in \eqref{Hermite} being those generated by attaching propagators exclusively to $\vp^\alpha$. (It corresponds to choosing $\ff^{\sigma}(\vpi,\mu) \to 2\pi A_\sigma\, i^\alpha \delta^{(\alpha)}(\vpi)$ as $\Lambda_\p\cu\to\infty$ \cite{first}.)

Now that the coefficient functions are polynomial, the whole linearised solution \eqref{complementary} is a polynomial. In particular it is now a sum over polynomial eigenoperators, where the latter 
are given by the $\Lambda$-independent $\sigma \vp^\alpha$ together with its finite number of $\Lambda$-dependent tadpole corrections generated by the exponential operator in \eqref{complementary}. The solutions are therefore effectively now also polynomial in $\hbar$, its power being given by the loop-order of these tadpole corrections. They are effectively no different from the solutions to the linearised flow equation \eqref{flowone} that we would write down in standard quantisation.

%\subsection{Solutions of the linearised mST}
%\label{sec:linearmST}

The second equation, the first-order mST \eqref{mSTone}, 
says that a linearised solution must be closed under the total free quantum BRST charge $\hs_0$. In the new quantisation BRST invariance is recovered only at scales much less than $\Lambda_\p$, where we enter the the diffeomorphism invariant subspace thanks to the trivialisations above.
%the amplitude suppression scale, $\Lambda_\p$, which is the cross-over point in fig. \ref{fig:flow} where the renormalized trajectory enters the diffeomorphism invariant subspace. 
%Equivalently BRST invariance is recovered in the limit $\Lambda_\p\cu\to\infty$. 
In particular at first order we have that 
\be 
\label{Gammaonelimit}
\Gamma_1\to \kappa\,(\cG_1+\cG_{1\q1})\,,\qquad \text{as}\quad \Lambda_\p\to\infty\,,
%\Gamma^2_1 \to \kappa\,\cG^2_1\,, \qquad \Gamma^1_1\to\kappa\,\cG^1_1\,,\qquad\Gamma^0_1\to \kappa\,\cG^0_1+\half\kappa\, b\Lambda^4\vp\,,\qquad \text{as}\quad \Lambda_\p\to\infty\,.
\ee
where $A_\sigma$ has been set to $\kappa$, and $\cG_1+\cG_{1\q1}$ is the free total quantum BRST cohomology representative, \ie is closed under $\hs_0$ but not exact. Here we take the opportunity to split it into the classical three-point vertex, $\cG_1$, and its one-loop tadpole correction,\footnote{$q$ stands for one-loop and the reason for the trailing 1 will become clear in the next section.} $\cG_{1\q1}$. It is the quantum correction needed to make the RHS a polynomial solution of the linearised flow equation, consistently with the required limit for the solution $\Gamma_1$. Note that $\kappa$ is thus to be viewed as an effective coupling which arises as a collective effect of all the underlying couplings, and which appears only in this $\Lambda,\vp\cu\ll\Lambda_\p$ regime.
%large amplitude suppression scale limit.  

In order to get a theory that is consistent with unitarity and causality, we restrict $\cG_1$ to have a maximum of two space-time derivatives. Then $\cG_1$ must be a linear combination of a term involving space-time derivatives and a unique non-derivative piece. 
%\be 
%\label{Gonecc}
%\cG_1 = \cG^0_1 = \vp\,.
%\ee
This latter is just $\vp$ itself, and is nothing but the $O(\kappa)$ part of $\sqrt{g}$, as we will review shortly. In this paper we set this first order cosmological constant term to zero. 
Up to an $\hs_0$-exact piece, the derivative part also has a unique expression  \cite{\morrii,Boulanger:2000rq}. We will use the choice  
%such that its antighost levels are given by 
\cite{first}
\beal 
\label{Gsplit}
\cG_1\ &=\ \cG^2_{1} + \cG^1_{1} + \cG^0_{1}\,,  \\
\label{Gonetwo}
\cG^2_{1}\ &=\ -\left( c^\mu\partial_\mu c^\nu\right) c^*_\nu \,,\\
\label{Goneone}
\cG^1_{1}\ &=\  - \left(c^\alpha \partial_\alpha H_{\mu\nu} + 2\, \partial_\mu c^\alpha h_{\alpha\nu}\right) H^*_{\mu\nu} -\vp\,\partial_\mu c_\nu H^*_{\mu\nu} \,,\\
\cG^0_{1}\ 
& = \tfrac14h_{\alpha\beta}\partial_\alpha\vp\partial_\beta\vp
-h_{\alpha\beta}\partial_\gamma h_{\gamma\alpha}\partial_\beta\vp
-\tfrac12 h_{\gamma\delta}\partial_\gamma h_{\alpha\beta}\partial_\delta h_{\alpha\beta}
-h_{\beta\mu}\partial_\gamma h_{\alpha\beta}\partial_\gamma h_{\alpha\mu}\nn \\
&\quad +2h_{\mu\alpha}\partial_\gamma h_{\alpha\beta}\partial_\mu h_{\beta\gamma} 
+h_{\beta\mu}\partial_\gamma h_{\alpha\beta}\partial_\alpha h_{\gamma\mu}
-h_{\alpha\beta}\partial_\gamma h_{\alpha\beta} \partial_\mu h_{\mu\gamma}
+\tfrac12h_{\alpha\beta}\partial_\gamma h_{\alpha\beta} \partial_\gamma \vp\nn \\
&\quad +\vp\left(\, \tfrac38(\partial_\alpha\vp)^2
-\tfrac12\partial_\beta h_{\beta\alpha}\partial_\alpha\vp
-\tfrac14(\partial_\gamma h_{\alpha\beta})^2 
+\tfrac12 \partial_\gamma h_{\alpha\beta}\partial_\alpha h_{\gamma\beta}\,\right)
%+\tfrac72 b \Lambda^4 \,\right)
\label{Gonezero} 
\eeal
Here we have split (graded) $\cG_1$ by antighost number (the superscript). The one-loop quantum part,
\be 
\label{Goneq}
\cG_{1\q1} = \cG^0_{1\q1} = \tfrac72 b \Lambda^4 \vp\,,
\ee
only has antighost level zero. The tadpole integral is written in terms of the non-universal dimensionless number \cite{Morris:2018mhd,\morrii,first}:
\be 
\label{b}
b = \int\!\frac{d^4\tilde{p}}{(2\pi)^4}\, C(\tilde{p}^2) =\frac1{(4\pi)^2}\int^\infty_0\!\!\!\!\!\!du\,u\, C(u)\,.
\ee
%being a non-universal dimensionless number \cite{Morris:2018mhd,\morrii,first} generated by this tadpole integral. 
%Although $\cG_{1\q1}$ satisfies \eqref{mSTone} on its own, and thus can be rescaled or discarded while still ensuring that $\cG_1$ satisfies \eqref{mSTone}, it is fixed uniquely by \eqref{flowone}.
 Although this reintroduces a first-order cosmological constant term, it is not physical since it vanishes in the limit $\Lambda\cu\to0$.

We will sometimes need $\cG_1$ in $d$ dimensions. In this case $\cG^2_1$ takes the same form, $\cG^1_1$ differs only in that $\vp$ should be replaced by $\frac4d\vp$ or alternatively this last term is removed and $h_{\alpha\nu}$ replaced by $H_{\alpha\nu}$. Finally, the $d$-dimensional level zero part is \cite{\morrii,first} 
\beal 
\cG^0_1\ 
& =\ 2 \vp\partial_\beta H_{\beta\alpha}\partial_\alpha\vp 
-2\vp(\partial_\alpha\vp)^2
-2H_{\alpha\beta}\partial_\gamma H_{\gamma\alpha}\partial_\beta\vp
+2H_{\alpha\beta}\partial_\alpha\vp\partial_\beta\vp
-2H_{\beta\gamma}\partial_\gamma H_{\alpha\beta}\partial_\alpha\vp\nn \\
&\phantom{=\ } +\half\vp (\partial_\gamma H_{\alpha\beta})^2 
-\half H_{\gamma\delta}\partial_\gamma H_{\alpha\beta}\partial_\delta H_{\alpha\beta}
-H_{\beta\mu}\partial_\gamma H_{\alpha\beta}\partial_\gamma H_{\alpha\mu}
+2H_{\mu\alpha}\partial_\gamma H_{\alpha\beta}\partial_\mu H_{\beta\gamma}\label{Gonezerod} \\
&\phantom{=\ }+H_{\beta\mu}\partial_\gamma H_{\alpha\beta}\partial_\alpha H_{\gamma\mu} 
-\vp \partial_\gamma H_{\alpha\beta}\partial_\alpha H_{\gamma\beta} 
-H_{\alpha\beta}\partial_\gamma H_{\alpha\beta} \partial_\mu H_{\mu\gamma}
+2H_{\alpha\beta}\partial_\gamma H_{\alpha\beta} \partial_\gamma \vp\,.\nn
\eeal

\section{Solving the Classical Master Equation}
\label{sec:solCME}

At intermediate steps we will need 
\be 
\label{cGkappaexp}
\cG = \sum_{n=0}^\infty \cG_n\, \kappa^n/n!\,,\qquad (\cG_0 = \Gamma_0)
\ee
where $\cG$ is a local solution of the \emph{Classical} Master Equation (CME), taking the standard  form:
\be 
\label{cGtot}
\cG = \cG^0 - (Q\Phi^A)\Phi^*_A\,.
\ee
In particular $\cG^0$ and $Q$ then also have such expansions in $\kappa$.
The CME \cite{ZinnJustin:2002ru,Batalin:1981jr,Batalin:1984jr}
\be 
\label{BRSTinv}
0 = \frac12 (\cG,\cG) = (Q\Phi^A) \frac{\partial_l \cG}{\partial\Phi^A}\,,
\ee
just implies the BRST invariance of this action under this \emph{classical} BRST charge $Q$. The choice of free total quantum BRST cohomology representative (\ref{Gonetwo},\ref{Goneone},\ref{Gonezero}) was made \cite{first} because $Q$ is then given exactly, \ie has no higher order in $\kappa$ corrections, provided that the metric is given by the simple linear split \eqref{gH}.
Indeed using the classical form \eqref{cGtot}, we read from $\cG^2_1$ \eqref{Gonetwo} that
\be 
\label{Qc}
Qc^\nu = (Q_0+\kappa\, Q_1)\, c^\nu = \kappa\, c^\mu\partial_\mu c^\nu = \half\,\kappa\, \mathfrak{L}_c\, c^\nu\,,
\ee
expresses exactly the algebra of diffeomorphisms through the Lie derivative $\mathfrak{L}_c$ generated by the vector field $\kappa c^\mu$ \cite{\morrii}, while from $\cG^1_1$ \eqref{Goneone} and the $H^*_{\mu\nu}$ part in  $\Gamma_0$ \eqref{Gzero} we get exactly the action of diffeomorphisms on the metric, through its Lie derivative:
\be 
\label{Qg}
Q g_{\mu\nu} = \kappa (Q_0 +\kappa Q_1) H_{\mu\nu} =  2\kappa\,\partial_{(\mu} c^\alpha g_{\nu)\alpha} + \kappa\, c^\alpha\partial_\alpha g_{\mu\nu} = \kappa\, \mathfrak{L}_c\, g_{\mu\nu}\,.
\ee
In our case, the level zero action, $\cG^0_1 +\cG^0_{1\q}$, has a classical and one-loop part. Together they must still solve these equations, indeed the CME and Zinn-Justin identities \cite{ZinnJustin:2002ru} are equivalent algebraically. Thus the one-loop part has an expansion in $\kappa$ which we write similarly to that for $\cG$ itself \eqref{cGkappaexp}:
\be 
\label{cGqkappaexp}
\cG^0_{1\q} = \sum_{n=1}^\infty \cG^0_{1\q n}\, \kappa^n/n!\,,
\ee
where the $O(\kappa)$ part is $\cG^0_{1\q1}$ as already given in \eqref{Goneq} (and the above now explains the notation). Since this quantum piece is purely level-zero it does not disturb the classical parametrisation \eqref{cGtot} and thus by the Zinn-Justin identities \eqref{BRSTinv} we now get the one-loop identity
\be 
\label{BRSTinvq}
0 = (\cG,\cG^0_{1\q}) = (Q\Phi^A) \frac{\partial_l \cG^0_{1\q}}{\partial\Phi^A}\,.
\ee
Using the (now-extended) classical form \eqref{cGtot} the identities \eqref{BRSTinv} follow from nilpotency, $Q^2\cu=0$, and the diffeomorphism invariance of $\cG^0$, while \eqref{BRSTinvq} expresses the diffeomorphism invariance of $\cG^0_{1\q}$. Expanding out the antibracket in \eqref{BRSTinv} to $O(\kappa^2)$ using the $\kappa$ expansion of the action \eqref{cGkappaexp}, the absence of classical corrections to $Q$ (\ref{Qc},\ref{Qg}) implies the first two of the following relations (which are readily verified), while the last two relations express the diffeomorphism invariance of $\cG^0$ and $\cG^0_{1\q}$ at second order:
% Expanding out the antibracket, the former is equivalent to the first two of the following relations (which are readily verified):
\be 
\label{CMErelations}
(\cG^2_1,\cG^2_1) = 0\,,\quad 2\,(\cG^2_1,\cG^1_1)+(\cG^1_1,\cG^1_1) = 0\,,\quad Q_0\,\cG^0_2 = -  (\cG^1_1,\cG^0_1)\,,\quad Q_0\,\cG^0_{1\q2} = -  (\cG^1_1,\cG^0_{1\q1})\,.
\ee
%while the last relation . 
Given that $\cG^0_{1\q1}$ \eqref{Goneq} is a $\Lambda$-dependent cosmological constant term  expanded to first order in $\kappa$, while the action for free gravitons \eqref{Gzero} covariantizes to the Einstein-Hilbert action for which $\cG^0_1$ \eqref{Gonezero} is its first order vertex \cite{\morrii}, 
%the sum at first-order of the vertex one gets from expanding the Einstein-Hilbert action using \eqref}  and , 
we know geometrically that all-orders solutions are
\be 
\label{geom}
%\,,\qquad\text{where}\quad 
\cG^0 = -2\sqrt{g}R/\kappa^2\,,\qquad \cG^0_{1\q}=\tfrac72b\Lambda^4\sqrt{g}\,,
\ee
where $R$ is the scalar curvature. Expanding \eqref{geom} to $O(\kappa^2)$ we thus find solutions  to the last equations in \eqref{CMErelations}, namely 
\besp
\label{Gbzerotwozero}
\cG^0_{2} = \vp^2 \Big(\tfrac14\partial_{\alpha}h_{\alpha \beta} \partial_{\beta} \vp - \tfrac3{16} (\partial_{\alpha} \vp)^2  
+ \tfrac{1}{8} (\partial_{\sigma} h_{\alpha \beta})^2  
- \tfrac{1}{4} \partial_\alpha h_{\beta\sigma}\partial_\beta h_{\alpha\sigma}\Big)
+\vp \Big(h_{\alpha \beta} \partial_{\sigma} h_{\sigma \alpha} \partial_{\beta} \vp  
\\
-\tfrac14\partial_\mu h^2_{\alpha\beta}\partial_\mu\vp
-\tfrac14h_{\alpha\beta}\partial_\alpha\vp\partial_\beta\vp
+\partial_\alpha h_{\alpha\beta}h_{\mu\nu}\partial_\beta h_{\mu\nu}
+\tfrac12\partial_\alpha h_{\mu\nu}\partial_\beta h_{\mu\nu} h_{\alpha\beta}
-2\partial_\mu h_{\nu\alpha} \partial_\beta h_{\mu\nu} h_{\alpha\beta}
\\
+\partial_\mu h_{\nu\alpha}\partial_\mu h_{\nu\beta} h_{\alpha\beta}
-\partial_\mu h_{\nu\alpha}\partial_\nu h_{\mu\beta} h_{\alpha\beta}\Big)
+\tfrac12\partial_\sigma h_{\sigma\alpha} h_{\alpha\beta} \partial_\beta h^2_{\mu\nu}
+\partial_\sigma h_{\sigma\alpha} \partial_\alpha h_{\beta\mu} h_{\beta\nu}h_{\mu\nu}
+\tfrac14\partial_\sigma h_{\sigma\alpha}\partial_\alpha\vp h^2_{\mu\nu}
\\
-\tfrac18 (\partial_\sigma h_{\alpha\beta})^2 h^2_{\mu\nu}
+\tfrac12\partial_\mu h_{\alpha\beta}\partial_\nu h_{\alpha\beta} h_{\mu\sigma} h_{\nu\sigma}
+\partial_\alpha h_{\beta\mu}\partial_\alpha h_{\beta\nu} h_{\mu\sigma} h_{\nu\sigma}
+\partial_\alpha h_{\sigma\mu} \partial_\beta h_{\sigma\nu} h_{\alpha\beta} h_{\mu\nu}
\\
-\partial_\alpha h_{\sigma\mu} \partial_\nu h_{\sigma\beta} h_{\alpha\beta} h_{\mu\nu}
-2\partial_\alpha h_{\beta\mu} \partial_\nu h_{\alpha\beta} h_{\mu\sigma} h_{\nu\sigma}
-\tfrac32 \partial_\mu h_{\nu\sigma}\partial_\sigma h_{\alpha\beta} h_{\alpha\mu} h_{\beta\nu}
+\tfrac12 \partial_\sigma h_{\alpha\beta} \partial_\sigma h_{\mu\nu}h_{\alpha\mu} h_{\beta\nu}
\\
+\tfrac14\partial_\sigma h_{\alpha\beta} \partial_\alpha h_{\sigma\beta} h^2_{\mu\nu}
+\tfrac12h_{\alpha\beta}\partial_\alpha h_{\beta\sigma} \partial_\sigma h^2_{\mu\nu}
-\partial_\alpha \vp \partial_\mu h_{\nu\alpha} h_{\mu\sigma} h_{\nu\sigma}
-\partial_\alpha h_{\beta\mu} \partial_\beta h_{\alpha\nu} h_{\mu\sigma} h_{\nu\sigma}
\\
-\tfrac12\partial_\alpha h_{\mu\sigma} \partial_\sigma h_{\beta\nu} h_{\alpha\beta} h_{\mu\nu}
+\partial_\alpha h_{\alpha\mu} \partial_\nu\vp h_{\mu\sigma} h_{\nu\sigma}
-\tfrac18(\partial_\mu h^2_{\alpha\beta})^2
-\tfrac3{16}h^2_{\mu\nu}(\partial_\alpha\vp)^2\,,
\eesp
which would be awkward to derive working directly with \eqref{CMErelations}, and 
\be \label{Gbonetwozero} \cG^0_{1\q2} = \tfrac78b\Lambda^4(\vp^2-h^2_{\alpha\beta}) \,. \ee
Note that other all-orders solutions are possible but will differ from \eqref{geom} by addition of further invariants at higher order in $\kappa$. At $O(\kappa^2)$ this is precisely the freedom we see in \eqref{CMErelations} to add $Q_0$-closed terms $\delta\cG^0_2$, which are thus also $\hs_0$-closed, \ie solutions to the linearised mST \eqref{mSTone}. 
These latter are explored further in sec. \ref{sec:BRSTexact}.

\section{BRST exact operators}
\label{sec:BRSTexact}

%Using also \eqref{KTHc} and \eqref{measure},
%\be 
%%\label{soeigenoperator}
%\hs_0\left( H_{\mu\nu} c_\mu c^*_\nu \right) = (\partial_\mu c_\nu + \partial_\nu c_\mu)\, c_\mu c^*_\nu +2H_{\mu\nu} c_\mu \partial_\alpha H^*_{\alpha\nu} + 2b\Lambda^4\vp\,,
%\ee 
%where we note that $\Delta^-$ trivially annihilates, but $\Delta^=$ yields a UV regulated quartically divergent contribution, $b$ being the non-universal number \eqref{b}.
%already introduced in refs. \cite{Morris:2018mhd,\morrii}:
%\be 
%\label{b}
%b = \int\!\frac{d^4\tilde{p}}{(2\pi)^4}\, C(\tilde{p}^2)\,.
%\ee

%\TRM{spell out three-point solutions including just renormalizing $\cG_1$?? Couldn't \eqrf{soeigenoperator} arise in quantum corrections? Not at one loop I think.}

%We have thus written our new choice for the two-derivative representative $\cG_1$ as the previous choice \cite{\morrii} plus the (quasi)local $\hs_0$-exact term \eqrf{soeigenoperator}.

We will see that at second order in perturbation theory \eqref{expansion}, local $\hs_0$-closed \emph{bilinear} terms, 
\be 
\label{closedbilinear}
\hs_0\,\delta \cG_2 = 0\,, 
\ee
play an important r\^ole (\cf sec. \ref{sec:standard}).
They appear with up to a maximum of four space-time derivatives and as we show now, turn out also to be $\hs_0$-exact. Such $\hs_0$-exact terms just reparametrise the free action and therefore carry no new physics \cite{Gomis:1994he,\morrii}. Indeed if we add an operator $\hs_0 K_2$ % = (Q_0+Q^-_0-\Delta) K_n$ 
to $\Gamma_0$ then, from the definitions of the free charges and linearised mST (\ref{charges},\ref{mSTone}) and the form of the antibracket \eqref{QMEbitsPhi}, we see that this corresponds to infinitesimal field and source redefinitions: 
\be 
\label{canon}
\delta \Phi^A = \frac{\partial_l K_2}{\partial \Phi^*_A}\,,\qquad \delta\Phi^*_A = -\frac{\partial_lK_2}{\partial\Phi^A}\,,
\ee
with the $-\Delta K_2$ part corresponding to the Jacobian of the change of variables in the partition function \cite{Batalin:1981jr,Batalin:1984jr}, regularised by $C^\Lambda$ \cite{\morrii,\yuji}.\footnote{In general it is exact expressions using the \emph{interacting} total BRST charge that correspond to  infinitessimal reparametrisations, however since we are interested only in changes at second-order and we are working at this order, only $\Gamma_0$ contributes, and not $\Gamma_1$.}

Since these local $\hs_0$-closed bilinear terms turn out also to be $\hs_0$-exact,
any $\mu$-dependence that they carry, can be eliminated by reparametrisation. This result is the BRST cohomological equivalent of the kinematical accident that pure gravity (without cosmological constant) is one-loop finite in standard quantisation \cite{tHooft:1974toh}, as we will highlight later.

Consider first the following two $\hs_0$-exact solutions:
\be 
\label{wavefunctionops}
\half \hs_0 (H^*_{\mu\nu} H_{\mu\nu}) = \partial_\mu c_\nu H^*_{\mu\nu} - H_{\mu\nu} G^{(1)}_{\mu\nu}\,,\qquad \half \hs_0(c^*_\mu c_\mu) = - \partial_\mu c_\nu H^*_{\mu\nu}\,,
\ee
where we used again the formula for $\hs_0$ \eqref{mSTone}, and the explicit actions for the charges   (\ref{QH},\ref{KTHc}) and always discard field independent terms.  The last term in the first equation is evidently again the action for free gravitons, while the remaining terms are up to a factor the source term in $\Gamma_0$ \eqref{Gzero}. These solutions generate the second order part of wavefunction renormalization $Z_E = 1+z_E$ ($E=H,c$), in close correspondence to the case of Yang-Mills \cite{\yuji}:
\be 
K_2 = \half z_H H^*_{\mu\nu} H_{\mu\nu} + \half z_c\, c^*_\mu c_\mu\,,
\ee
the full wavefunction renormalization being given by the finite (classical) canonical transformation 
\be 
K = \sum_E Z^{\half}_E\Phi^*_E\Phi^E_{(r)}\,, \qquad 
\Phi^E = \frac{\partial_l}{\partial\Phi^*_E} {K}[\Phi_{(r)},\Phi^*]\,,\qquad \Phi^*_{(r) E} = \frac{\partial_r}{\partial \Phi^E_{(r)}} {K}[\Phi_{(r)},\Phi^*]\,,
\ee
the subscript $(r)$ labelling the renormalized (anti)fields. This implies that the fields and antifields renormalize in opposite directions:
\be
H_{\mu\nu} = Z^{\half}_H H_{(r)\mu\nu}\,,\ H^*_{\mu\nu} = Z^{-\half}_H H^*_{(r)\mu\nu}\,,\ c_\mu = Z^{\half}_c c_{(r)\mu}
\,,\ c^*_\mu = Z^{-\half}_c c^*_{(r)\mu}\,.
\ee
However here this is not the whole story, in particular because the reparametrisations that are generated by quantum corrections are more general than this. 

Returning to the annihilation condition \eqref{closedbilinear}, we note that since $\delta\cG_2$ is bilinear and must have ghost number zero overall, it cannot have antighost number larger than one. At lowest order in derivatives, there are only two linearly independent possibilities for the $\delta\cG^1_2$ part, namely $H^*_{\mu\mu}\partial_\alpha c_\alpha$ and $H^*_{\mu\nu}\partial_\mu c_\nu$. The latter option solves \eqref{closedbilinear} since it is $\hs_0$-exact; it was treated already  \eqref{wavefunctionops}. By inspection \eqref{QH}, the former is $Q_0$-exact, and thus we know that  
it completes to an $\hs_0$-exact solution
\be 
\label{wavefunctionphi}
\hs_0(\vp^*\vp) = \vp^*\partial\cu\cdot c - R^{(1)}\vp\,,
\ee
where we have also split the graviton antighost into its $SO(4)$ irreducible parts:
\be 
H^*_{\mu\nu} = h^*_{\mu\nu}+\half \vp^*\delta_{\mu\nu}\,, \qquad \vp^* = \half H^*_{\mu\mu}\,, 
\ee
and recalled the standard relation for $G_{\mu\nu}$ \eqref{Gmunu}. Comparing to the structure in the previous paragraph, it is evident that \eqref{wavefunctionphi} expresses the fact that the $SO(4)$ irreducible parts can have separate wavefunction renormalizations. The remaining possibility at second order in derivatives, is to have a separate $\delta\cG^0_2$ part, but for it to be annihilated by $\hs_0$ \eqref{closedbilinear} it must be invariant under linearised diffeomorphisms \eqref{QH} and the graviton action in \eqref{Gzero} is the unique such solution at this order in derivatives. Any change in the graviton action normalization is already taken care of by a canonical transformation, being a linear combination of the two $\hs_0$-exact operators in \eqref{wavefunctionops}.

At next order in derivatives there are three linearly independent possibilities for $\delta \cG^1_2$, namely $\vp^*\Box\partial\cu\cdot c$, $H^*_{\mu\nu}\Box \partial_\mu c_\nu$,  and $H^*_{\mu\nu} \partial^3_{\mu\nu\alpha} c_\alpha$. Evidently the first yields a simple generalisation of $\vp$ wavefunction renormalization \eqref{wavefunctionphi}, while the second two are already $\hs_0$-exact:
\be
\label{fourthorderexact}
\hs_0(\vp^*\Box\vp) = \vp^*\Box\partial\cu\cdot c - R^{(1)}\Box\vp\,,\ \,\half \hs_0(c^*_\mu \Box c_\mu) = H^*_{\mu\nu}\Box \partial_\mu c_\nu\,,\quad -\hs_0(H^*_{\mu\nu}\partial^2_{\mu\nu}\vp) = H^*_{\mu\nu} \partial^3_{\mu\nu\alpha} c_\alpha\,.
\ee 
The remaining possibility is to have a separate $\delta\cG^0_2$ part, now fourth-order in derivatives. Since it must be invariant under linearised diffeomorphisms, it has to be a linear combination of the squares of the linearised curvatures \eqref{curvatures} (see \eg \cite{Morris:2016nda}). By the  Gauss-Bonnet identity, only two of these are linearly independent:
\be 
\label{GB}
4(R^{(1)}_{\mu\nu})^2 = (R^{(1)}_{\mu\nu\alpha\beta})^2+(R^{(1)})^2\,.
\ee
However it is straightforward to see that they are also $\hs_0$-exact:
\beal 
\hs_0(\vp^*R^{(1)}) &= Q^-_0(\vp^*R^{(1)}) = -(R^{(1)})^2\,,\nn\\ 
\label{exactinvariant}
\hs_0(H^*_{\mu\nu} R^{(1)}_{\mu\nu}) &= -2 G^{(1)}_{\mu\nu} R^{(1)}_{\mu\nu} \,= \half (R^{(1)}_{\mu\nu\alpha\beta})^2 -\half(R^{(1)})^2\,.
\eeal
This completes the demonstration that the $\hs_0$-cohomology of  bilinear $\delta\cG_2$ is trivial up to the fourth order in derivatives.
%We have therefore shown that all solutions to \eqref{closedbilinear} up to the fourth order in derivatives, just correspond to reparametrisations of the action at this order in perturbation theory.\footnote{Since these induce second order changes, only $\Gamma_0$ is involved (not $\Gamma_1$).}

We note in passing that there are other expressions for the $\hs_0$-exact operators, for example the obvious generalisation of the first equation in wavefunction reparametrisations  \eqref{wavefunctionops}: 
\be 
\label{notanotheroption}
\half \hs_0 (H^*_{\mu\nu} \Box H_{\mu\nu}) = \partial_\mu c_\nu \Box H^*_{\mu\nu} - H_{\mu\nu} \Box G^{(1)}_{\mu\nu} =  \partial_\mu c_\nu \Box H^*_{\mu\nu}+  \half(R^{(1)})^2 -\half (R^{(1)}_{\mu\nu\alpha\beta})^2\,.
\ee
However these are not linearly independent, \eg the above
is a linear combination of the second exact expression in \eqref{exactinvariant} and the middle one in \eqref{fourthorderexact}. Stated another way, we have shown that the following action functional is annihilated by $\hs_0$: 
\be 
K_2 = \half H^*_{\mu\nu} \Box H_{\mu\nu} + \half c^*_\mu\Box c_\mu + H^*_{\mu\nu}R^{(1)}_{\mu\nu} = \half \hs_0(c^*_\mu F_\mu)\,.
\ee
This is so because in fact it itself is exact. The appearance of the De Donder gauge fixing functional,
$F_\mu = \partial_\nu H_{\nu\mu} -\partial_\mu\vp$, is here accidental. The most general double-derivative bilinear $\hs_0$-exact $K_2$ is a linear combination involving the two separate parts of $F_\mu$: 
\be 
\label{Ktwo}
K_2 = \hs_0( \alpha c^*_\mu \partial_\mu \vp + \beta c^*_\mu \partial_\nu H_{\mu\nu} )
= \alpha ( 2H^*_{\mu\nu}\partial^2_{\mu\nu}\vp +c^*_\mu \partial^2_{\mu\nu} c_\nu ) +\beta (H^*_{\mu\nu} \partial^2_{\mu\lambda} H_{\lambda\nu}+c^*_\mu \partial^2_{\mu\nu} c_\nu + c^*_\mu\Box c_\mu)\,,
\ee
where $\alpha$ and $\beta$ are free parameters.
The chosen $K_2$\,s in \eqref{fourthorderexact} are only $\hs_0$-cohomology representatives determined up to addition of the above expression. Since the above expression is annihilated by $\hs_0$, the canonical transformation \eqref{canon} it generates is actually a (higher-derivative) symmetry of $\Gamma_0$:
\beal 
\delta H_{\mu\nu} = 2\partial_{(\mu} \xi_{\nu)}\,,\qquad \text{where}\quad \xi_\mu &= \alpha\partial_\mu\vp+\beta\partial_\lambda H_{\lambda\nu}\,,\nn\\
\delta c_\mu =(\alpha+\beta)\partial^2_{\mu\nu}c_\nu+\beta\Box c_\mu\,,\qquad \delta H^*_{\mu\nu} &= -\alpha\delta_{\mu\nu} \partial^2_{\alpha\beta} H^*_{\alpha\beta} -2\beta\partial_\alpha \partial_{(\mu} H^*_{\nu)\alpha}\,.
\eeal
For the graviton it is just part of linearised diffeomorphism invariance. The significance of the other two transformations is unclear to us. They may not survive into the interacting theory.

\section{Inside the diffeomorphism invariant subspace}
\label{sec:evaluation}

At second order in perturbation theory \eqref{expansion}, the flow equation \eqref{flow},  and mST \eqref{mST}, become
\beal 
\label{flowtwo}
\dot{\Gamma}_2 - \half\, \text{Str}\, \dot{\prop}_\Lambda \Gamma^{(2)}_2 &= 
- \half\, \text{Str}\, \dot{\prop}_\Lambda \Gamma^{(2)}_1 \propH \Gamma^{(2)}_1 \,, \\
\label{mSTtwo}
 \hs_0\,\Gamma_2 &= -  \half\,(\Gamma_1,\Gamma_1) -\text{Tr}\,C^\Lambda\, \Gamma^{(2)}_{1*}\propH \Gamma^{(2)}_1 \,.
%\left( \!C_\Lambda\,  \Gamma^{(2)}_{1*}\propH \Gamma^{(2)}_1  \right) \,.
\eeal
In ref. \cite{secondconf} we constructed the general continuum limit solution to \eqref{flowtwo}, \ie the general solution that realises the full renormalized trajectory $\Lambda\cu\ge0$. It takes the form
\be 
\label{Gammatwosol}
\Gamma_{2} = \half \left[\, 1+\Po_\Lambda-(1+\Po_\mu)\,\mathrm{e}^{\Po^\mu_\Lambda}\,\right] \Gamma_{1}\, \Gamma_{1} + \Gamma_{2}(\mu)\,,
\ee
where the first term on the RHS is the particular integral and the last term is the complementary solution. The complementary solution takes exactly the form of the general solution  \eqref{complementary} to the linearised flow equation, where however $\mu$ now has a meaning. It is an arbitrary initial point on the renormalized trajectory, lying in the range $0<\cu\mu\cu < a\Lambda_\p$. The particular integral expands out into a sum over melonic Feynman diagrams, the propagators defined through (similarly $\Po_\mu$ and $\Po_\Lambda$)
\be 
\Po^\mu_\Lambda = {\prop}^{\mu\,AB}_\Lambda \frac{\partial^L_l}{\partial\Phi^B}\frac{\partial^R_l}{\partial\Phi^A} \,.
\ee
They connect the two copies of the first-order solution $\Gamma_1$. Importantly the renormalized trajectory solution \eqref{Gammatwosol} is already finite, all the UV divergences having been absorbed into the relevant underlying second-order couplings, $g^\sigma_{2l+\eps}$, as described in ref. \cite{secondconf}.

We now describe the properties of these equations and their solution once the renormalized trajectory has entered the diffeomorphism invariant subspace, \cf fig. \ref{fig:flow}. This is equivalent in particular to taking the large $\Lambda_\p$ limit. In sec. \ref{sec:standard} we then provide the detailed solution. 

In the large $\Lambda_\p$ limit, the limit at first order \eqref{Gammaonelimit} can be substituted directly into the second-order mST \eqref{mSTtwo} and into the particular integral, since these expressions are well defined being both regularised in the IR and the UV and 
and remain so in this limit \cite{secondconf}. Since $\cG_1$ contains a maximum of three fields, the latter then collapses to a one-loop integral in the sense that the renormalized trajectory \eqref{Gammatwosol} now reads \cite{secondconf}
\be 
\label{partLASS}
\Gamma_2 
= %\sum_\sigma \left( \sigma f^\sigma_\Lambda(\vp,\mu)+\cdots \right)
\Gamma_2(\mu) +\kappa^2\left( I_{2\Lambda} - I_{2\mu}\right)\,,\quad\text{where}\quad I_{2k} = 
-\tfrac14\,\text{Str}\!\left[ %\proph{\mu} \check{\Gamma}^{(2)}_1 \proph{\mu} \check{\Gamma}^{(2)}_1
\proph{k} \check{\Gamma}^{(2)}_1 \proph{k} \check{\Gamma}^{(2)}_1\right]\,,\quad (k=\mu,\Lambda)\,.
\ee
In fact $I_{2k}$ is now identical to a one-loop computation  in standard quantisation. Although  it is built from first-order vertices, which themselves contain a one-loop tadpole contribution $\cG_{1\q1}$  \eqref{Goneq}, this latter drops out because it is linear in $\vp$. In a similar way the RHS of the second-order mST \eqref{mSTtwo} can be seen to contain all the standard quantisation one-loop contributions and no more. 

At this stage the infinite number of underlying  couplings have disappeared, leaving behind only $\kappa$. Had we chosen to keep a first-order cosmological constant, then it would also appear as an effective coupling. As we will see, similarly to the standard perturbative approach, further effective couplings generically appear order by order in perturbation theory, multiplying covariant higher derivative terms (such as curvature squared terms \etc).
%a handful of effective couplings, one of which is $\kappa$.  
Here however these effective couplings are collective effects of the infinite number of underlying couplings, and parametrise the remaining freedom in the renormalized trajectory given that it has entered the diffeomorphism invariant subspace, \cf fig. \ref{fig:flow}. 
In our case from here on we can identify the perturbative expansion as being an expansion in $\kappa$. Therefore we redefine the second order contribution to be $\kappa^2\,\Gamma_2$ with complementary solution $\kappa^2\,\Gamma_2(\mu)$, so that from here on $\kappa$ drops out of the equations.

%In particular the complementary solution is now a sum over polynomial eigenoperators, where the latter 
%is given by the $\Lambda$-independent $\sigma \vp^\alpha$, and its $\Lambda$-dependent tadpole corrections generated by the exponential operator in \eqref{complementary}. Recall that the treatment should be non-perturbative in $\hbar$ \cite{\morri,\morrii,first,secondconf}. However since the field dependence is now polynomial, in practice the dependence on $\hbar$ can also be treated as polynomial (its degree being the maximum number of loops that can be created with such vertices at the given order in $\kappa$). 

The particular integral is  now polynomial in the fields. 
In this limit we also arrange for $\Gamma_2(\mu)$ to trivialise, \ie become polynomial, as explained in sec. \ref{sec:sollinear}. 
We see therefore that from a practical point of view the computation can now proceed in a way which is very close to standard quantisation. We comment further in the Conclusions. We emphasise that the understanding of the result is however very different: in standard quantisation, $\kappa$ is a fundamental irrelevant coupling and thus there is no interacting continuum limit in the Wilsonian sense \cite{\morri,Morris:2018upm}. Here the continuum limit is expressed in terms of the infinite number of underlying couplings, which are all (marginally) relevant. It is these latter that get renormalized in this picture, as noted above.
%\footnote{In \TRM{the Conclusions?}, we note the quantum field theory can be understood in a simpler and more radical way.}
%At higher orders in $\kappa$ it looks like the large-$\Lambda_\p$ limit might differ from standard quantisation even from a practical point of view, but only because not all contributions perturbative in $\hbar$ are reproduced up to the maximum number of loops that appear. It seems that this would just be a finite reordering of the loop-wise expansion, without physical consequence. 

%However in the current case we can see that the computation is in practice identical to a one-loop calculation in standard quantisation. The $O(\kappa)$ vertex contains already its corresponding one-loop (tadpole) contribution $\cG_{1\q1}$, \cf \eqref{Goneq}. The particular integral in \eqref{partLASS} is of one-loop form. Although it is built from first-order vertices, it is still maximum one-loop because $\cG_{1\q1}$ is linear in $\vp$ and thus drops out. For similar reasons the mST \eqref{mSTLASS} also contains all one-loop contributions and no more.

From the perspective of standard quantisation, the large-$\Lambda_\p$ limit  \eqref{partLASS} still looks a little peculiar since the particular integral is the difference of two parts: $I_{2\Lambda}\cu - I_{2\mu}$. These parts are IR regulated but separately UV divergent. 
%Now notice that the two parts of the particular integral (respec. $\mu$ and $\Lambda$ dependent) in \eqref{partLASS}  
We can treat them separately by applying some appropriate supplementary regularisation, \eg dimensional regularisation,
%\be 
%\label{dimreg}
$d=4-2\epsilon$,
as was done in ref. \cite{\yuji}. Furthermore we can subtract their divergences separately using a gauge invariant scheme that is independent of the finite cutoff scale $\mu$ or $\Lambda$, since such divergences anyway cancel out between the two parts. We will use the $\overline{\text{MS}}$ (modified minimal subtraction) scheme, and thus subtract the terms proportional to
%\be 
%\label{MSbar}
$1/\epsilon-\gamma_E+\ln(4\pi)$, where $\gamma_E$ is the Euler-Mascheroni constant.
%\,.
%\ee
%defining the Feynman integrals to be UV subtracted using some appropriate supplementary regularisation,  such that the subtraction is independent of the finite cutoff scale (\viz $\mu$ or $\Lambda$), as was done in ref. \cite{\yuji}. These subtractions then just cancel out between the two parts. 

Since $\cG_1$ is made of three-point vertices, the particular integral contains only two-point vertices.
%the $\prop_\mu$ part of \eqref{partLASS} contains only two-point vertices. 
When derivative-expanded, $I_{2\mu}$ trivially results in polynomial (in the fields) solutions to the linearised flow equation \eqref{flowone}, because these carry no $\Lambda$ dependence and the tadpole corrections, where they exist, are field independent and thus --although calculable-- discarded since they contain no physics. We can therefore dispense with $I_{2\mu}$ by absorbing it into a redefinition of the complementary solution: $\Gamma_2(\mu)\mapsto \Gamma_2(\mu)+I_{2\mu}$. As further discussed below, this is essentially what we will do except that we will take due account of the fact that $I_{2\Lambda}$ is ambiguous on its own, whereas in fact the difference that appears in the large-$\Lambda_\p$ limit \eqref{partLASS} is finite and well defined. %This is important because it is the latter property that leads to crucial differences in treatment compared to standard quantisation. 
%(We also note that this redefinition becomes more subtle when there are non-trivial tadpole corrections that could be attached to the $\mu$-part.

Emphasising the similarity to the standard perturbative approach we now write:
\beal 
\label{flowtwocl}
\Gamma_{2\cl}  &= \Gamma_{2\cl}(\mu) \\ 
\label{mSTtwocl}
s_0 \Gamma_{2\cl} &= - \half\,(\cG_1,\cG_1) \,,\\
\label{flowtwoq}
\Gamma_{2\q} &= \Gamma_{2\q}(\mu) +I_{2\Lambda} - I_{2\mu}\,,\\
%+\tfrac14\,\text{Str}\!\left[ \proph{\mu} \check{\Gamma}^{(2)}_1 \proph{\mu} \check{\Gamma}^{(2)}_1 - \proph{\Lambda} \check{\Gamma}^{(2)}_1 \proph{\Lambda} \check{\Gamma}^{(2)}_1\right]\,, \\
\label{mSTtwoq}
s_0 \Gamma_{2\q} -\Delta \Gamma_{2\cl} &= -(\cG_1,\cG_{1\q})-\text{Tr}\,C^\Lambda\, \cG^{(2)}_{1*}\propH \cG^{(2)}_1 \,.
\eeal
Here we have split the solution $\Gamma_2 = \Gamma_{2\cl}+\Gamma_{2\q}$  \eqref{partLASS} to the second-order flow equation, into its classical \eqref{flowtwocl} and one-loop \eqref{flowtwoq} parts, and similarly split the complementary solution: $\Gamma_{2}(\mu) = \Gamma_{2\cl}(\mu)+\Gamma_{2\q}(\mu)$. We have also split the second-order mST \eqref{mSTtwo} into its classical \eqref{mSTtwocl} and one-loop \eqref{mSTtwoq} parts, noting by definition of the total free quantum BRST charge \eqref{mSTone}, that $\hs_0 = s_0 -\Delta$, where $s_0=Q_0+Q^-_0$ is the classical part, while 
the measure operator $\Delta$ is $O(\hbar)$ \cite{Batalin:1981jr,Batalin:1984jr,\yuji}. 

%We have also split $\Gamma_2 = \Gamma_{2\cl}+\Gamma_{2\q}$ into its classical and one-loop parts, and similarly split the complementary solution: $\Gamma_{2}(\mu) = \Gamma_{2\cl}(\mu)+\Gamma_{2\q}(\mu)$.

%off the one-loop part, $\Gamma_{2\q}$, from the second-order solution, relabelling $\Gamma_2$ as the classical solution, and similarly split off from the complementary solution its one-loop part, $\Gamma_{2\q}(\mu)$. 

The trivialised complementary solution is just a polynomial (in fields)  solution to the linearised flow equation \eqref{flowone}, so $\Gamma_{2\cl}(\mu)$ is a $\Lambda$-independent part, while $\Gamma_{2\q}(\mu)$ contains the induced $\Lambda$-dependent one-loop tadpole correction plus its own $\Lambda$-independent part. In principle (and in general at higher order) there could be higher-loop tadpoles, however we will shortly see that in our case $\Gamma_2(\mu)$ only has a one-loop tadpole, while the one-loop $\Lambda$-independent part has no tadpoles. Therefore \eqref{flowtwocl}--\eqref{mSTtwoq} form the complete set of $O(\kappa^2)$ equations in our case. 

The classical flow equation \eqref{flowtwocl} simply says that $\Gamma_{2\cl}$ must be $\Lambda$-independent. 
%With this understanding it can be trivially satisfied by eliminating $\Gamma_2(\mu)$ in favour of $\Gamma_2$. 
If we absorb $I_{2\mu}$ entirely into $\Gamma_{2\q}(\mu)$ as discussed above,
the remaining three equations \eqref{mSTtwocl}--\eqref{mSTtwoq} are then identical to those we would derive in standard quantisation at one loop in this framework \cite{\yuji}. Given that we have defined $I_{2\Lambda}$
using dimensional regularisation and a gauge invariant subtraction scheme such as $\overline{\text{MS}}$, we then find a unique finite solution to these equations, up to the usual arbitrary  $\ln\mu_R$ terms appearing after subtracting logarithmic divergences, where the mass scale $\mu_R$ arises from dimensionally continued couplings (here $\kappa\mu_R^\epsilon$). The insertion of the cutoff $\Lambda$ leads to the modified Slavnov-Taylor identity \eqref{mSTtwoq}, but for vertices defined using a gauge invariant scheme such as $\overline{\text{MS}}$, this is still just an identity that is automatically satisfied. 
%(as we will confirm, see also \cite{\yuji}).

This is however a rather confusing way to arrive at a solution, because in our case  ambiguities such as the $\mu_R$-dependence cancel out in the difference $I_{2\Lambda}\cu - I_{2\mu}$, reflecting the fact that the quantum part of our solution \eqref{flowtwoq} is actually a well-defined expression. We have instead a mass parameter $\mu$ which plays essentially the same r\^ole, being the arbitrary initial point on the renormalized trajectory. Indeed like $\mu_R$ in the standard approach, physical quantities must ultimately be independent of $\mu$. We therefore choose to absorb all of $I_{2\mu}$ except essentially for exchanging $\mu_R$ with $\mu$. 
%which thus gives us a well defined complementary solution. 
As we will see $\overline{\text{MS}}$ then amounts to imposing a renormalization condition at $\mu\cu=\mu_R$, in the form expected in this framework \cite{\yuji}.

The failure point of standard perturbative quantisation is usually seen as stemming from the need to introduce bare couplings to absorb the UV divergences. Since in standard quantisation these multiply new non-trivial BRST cohomology representatives order by order in perturbation theory,  new bare couplings are needed at each order.
%\footnote{These include covariant higher derivative terms that are dangerous since they can destroy unitarity by introducing poles of the wrong sign into the propagator \cite{Stelle:1976gc}.}  
However we do not need direct access to the UV divergences to see the problem. The freedom to change the scheme away from $\overline{\text{MS}}$ to some other gauge invariant scheme, is contained in the freedom to add suitable local terms associated to the ambiguities in the finite parts of these divergences. \trm{The undetermined parameters that are thus required to parameterise the scheme dependence, are nothing but the new couplings that we know appear} at each order in standard quantisation. It is just that phrased this way the required new couplings are finite.
%in that we have to introduce new non-trivial BRST cohomology representatives order by order in perturbation theory, and thus new couplings at each order. 
Even if we stay within the $\overline{\text{MS}}$ scheme, $\mu_R$ independence would force the introduction of new finite couplings.

Here the UV divergences have already been absorbed into underlying (non-geometric) second order couplings $g^\sigma_{2l+\eps}$, and the ambiguities in defining the integrals are absent since they cancel out in the difference, $I_{2\Lambda}-I_{2\mu}$. Nevertheless there remains order by order in $\kappa$ the equivalent freedom. Indeed the requirement that our general second-order solution for the renormalized trajectory \eqref{Gammatwosol} %\footnote{or its specialised forms (\ref{partLASS},\ref{flowtwoq})}
 is independent of the initial point $\mu$,  will force the existence of the new effective couplings in the same way.\footnote{\label{foot:muindependence}Thus also its large-$\Lambda_\p$ limit  (\ref{partLASS},\ref{flowtwoq}). This is so in general even if inconveniently for us, for pure quantum gravity at $O(\kappa^2)$ such additions turn out to be $\hs_0$-exact, as we saw in sec. \ref{sec:BRSTexact}, and therefore can be removed by reparametrising the (anti)fields. As we noted in sec. \ref{sec:BRSTexact}, this is equivalent to the observations made in ref. \cite{tHooft:1974toh}.} More generally we have the freedom to add a local term to the solution $\Gamma_2$ of the second-order flow and mST equations (\ref{flowtwo},\ref{mSTtwo}), provided that
this addition satisfies just their left hand sides, \ie the linear equations (\ref{flowone},\ref{mSTone}). In other words it is a change in the complementary solution $\Gamma_2(\mu)$ corresponding to a change in our choice of (quantum) BRST cohomology representative. In particular once we have secured one solution for $\Gamma_2$ (\eg using the technique sketched above), we then have all possible solutions since they differ only by such a change in the quantum BRST cohomology representative. 
Since we already know that $I_{2\Lambda}$ on its own, defined with a suitable gauge invariant scheme, will satisfy the equations, we know that its scheme ambiguities are contained in such changes to the complementary solution.
% Similarly, we will see that the requirement that the solution \eqref{Gammatwosol}\footnote{or its specialised forms (\ref{partLASS},\ref{flowtwoq})} is independent of the initial point $\mu$,  then forces the existence of the new effective couplings in much the same way.\footnote{\label{foot:muindependence} This is so in general even if inconveniently for us, for pure quantum gravity at $O(\kappa^2)$ such additions turn out to be $\hs_0$-exact, as we saw in sec. \ref{sec:BRSTexact}, and therefore can be removed by reparametrising the (anti)fields. As we noted in sec. \ref{sec:BRSTexact}, this is equivalent to the observations made in ref. \cite{tHooft:1974toh}.}

We therefore have to confront the possibility that, although perturbatively in $\kappa$ we have a genuine continuum limit (at least to second order as confirmed here), it is of an unusual form in that the renormalized trajectory is parametrised by an infinite number of effective couplings.
%The dimensionally regularised integrals contain up to quartic UV divergences. 
%As usual, \old{the freedom to change the scheme away from $\overline{\text{MS}}$ is contained in the freedom to add local terms to these two-point vertices, associated to the ambiguities in the finite parts of these divergences.} However again such local terms cancel out in the difference. %that forms the particular integral. 
\textit{A priori} there seems to be nothing inconsistent with such a conclusion for quantum gravity, no matter how phenomenologically inconvenient,\footnote{In the general case, these include couplings for curvature-squared terms,  whose sign must be chosen to maintain unitarity, in contrast to the case
where quantum gravity would then be renormalizable in standard quantisation \cite{Stelle:1976gc}.} as we discuss further in sec. \ref{sec:discussion}. However in sec. \ref{sec:missing} we uncover hints that the non-polynomial dependence on $h_{\mu\nu}$ required by diffeomorphism invariance should force the BRST cohomology at the non-perturbative level back to be at most two-dimensional, depending only on $\kappa$ and the cosmological constant.

\subsection{Vertices at second order}
\label{sec:standard}

%Although we have arrived at a description equivalent to standard perturbative computations in quantum gravity, in this description the continuum action and BRST invariance is derived directly without first constructing bare versions. Since this framework \eqref{mSTtwocl}--\eqref{mSTtwoq} is unfamiliar \cite{\yuji}, we walk through the solution.
We now fill in the details.
%solve the equations \eqref{flowtwocl}--\eqref{mSTtwoq} in detail. 
We have already noted that \eqref{flowtwocl} just says that $\Gamma_{2\cl}$ is $\Lambda$-independent. 
From the first three equations \eqref{CMErelations} derived from the CME, it is clear that the choice we require so as to satisfy the  classical BRST invariance \eqref{mSTtwocl}, is 
\be 
\label{Gammatwocl}
\Gamma_{2\cl} = \Gamma_{2\cl}(\mu) = \cG^0_2\,.
\ee
It is therefore actually independent of $\mu$.
As anticipated, it only has a one-loop tadpole, 
%$\Gamma_{2\q}(\mu) = \cG^0_{2\q2}$, where
%and thus we find a contribution $\cG^0_{2\q2}\in\Gamma_{2\q}(\mu)$ where 
\be 
\label{comptad}
\Gamma_{2\q}(\mu) \ni \cG^0_{2\q2} = \Omega_\Lambda \Big(\tfrac32(\partial_\alpha\vp)^2-2\partial_\alpha h_{\alpha\beta}\partial_\beta\vp-(\partial_\sigma h_{\alpha\beta})^2+2(\partial_\alpha h_{\alpha\beta})^2
\Big) -\tfrac34b\Lambda^4(\vp^2+h^2_{\alpha\beta})\,,
\ee
%is the tadpole correction to \eqref{Gbzerotwozero} computed via \eqref{complementary}. 
computed using the classical $O(\kappa^2)$ expression \eqref{Gbzerotwozero} and the tadpole corrections defined in the general form of the complementary solution \eqref{complementary}, and labelled using the system introduced in \eqref{cGqkappaexp}.
(Notice that this involves the trivialisation of $\alpha\cu=2$ coefficient functions \eqref{flatp}, as is clear from the top line of the classical $O(\kappa^2)$ expression \eqref{Gbzerotwozero}, but their tadpole corrections are also joined by $h_{\mu\nu}$-tadpole corrections from the bottom lines in \eqref{Gbzerotwozero}.)\footnote{\textit{E.g.} $(\partial_\alpha\vp)^2$ arises from the second and the last monomial in \eqref{Gbzerotwozero} yielding, by \eqref{Hermite} and \eqref{hh}, $-\frac3{16}(1-9) =\frac32$.} %Also note that in \eqref{mSTtwoq}, we see in this case $\Delta\Gamma_2$ vanishes trivially.)
If we had already absorbed $I_{2\mu}$ into $\Gamma_{2\q}(\mu)$, \eqref{comptad} would actually be the complete solution for $\Gamma_{2\q}(\mu)$, being the unique $O(\kappa^2)$ tadpole integral formed from the classical action.

By inspection the particular integral \eqref{partLASS} and the RHS of the one-loop second-order mST \eqref{mSTtwoq} can contribute only up to a maximum antighost level two. In fact there is no contribution even at this level, as we now show. In the particular integral this would require attaching two propagators between $\cG^2_1$ \eqref{Gonetwo} and $\cG^1_1$ \eqref{Goneone}, or between two copies of $\cG^1_1$ while preserving both antifields, but it is not possible to attach the propagators in this way.
%it is not possible to attach two propagators between \eqref{Gonetwo} and \eqref{Goneone}, or between two copies of \eqref{Goneone} while preserving both antifields. 
Since $\cG_{1\q}$ only has level zero, the antibracket cannot contribute above level zero, whilst there is no correction term at level two in the one-loop second-order mST since this would require $\cG^{2\,(2)}_1$, but there is no way to join this by a propagator to $\cG^{(2)}_{1*}$. Thus all these antighost levels are solved by
%Since we need only one trivialisation that works, we therefore set all 
$\Gamma^{n\ge2}_{2\q} = \Gamma^{n\ge2}_{2\q}(\mu) = 0$.

For similar reasons the one-loop second-order mST \eqref{mSTtwoq} also collapses at antighost level one: 
\be 
\label{mSTtwoone}
Q_0\, \Gamma^1_{2\q} = 0\,,
\ee 
indeed the correction term now requires $\cG^{1(2)}_1$ with its antifield intact, but no such contributions are possible. However at this level the particular integral does make a contribution. The integral
\be 
\label{B}
I^1_{2\Lambda} =
i\! \int_p\!\! H^*_{\mu\nu}(p)\, \mathcal{B}^I_{\mu\nu\alpha}(p,\Lambda)\, c_\alpha(-p)
\ee 
%(using \eqref{defs}) 
is a two-point vertex formed from two copies of $\cG^1_1$ \eqref{Goneone} and fluctuation and ghost propagators \eqref{HH}--\eqref{cc} in the self-energy contribution \eqref{partLASS}. In $d\cu=4$ dimensions \com{ (1103.5)}
\besp
\label{Bint}
\mathcal{B}^I_{\mu\nu\alpha}(p,\Lambda) = -\int_q\!\frac{C_\Lambda(q)\,C_\Lambda(p\cu+q)}{q^2}\Big\{\frac{1}{(p\cu+q)^2}\left[{\tfrac32}^{\vphantom{a}}\,p_\alpha p_{(\mu} q_{\nu)}+\tfrac32\, p_\mu p_\nu q_\alpha+3\,p_{(\mu} q_{\nu)} q_\alpha+p^2p_{(\mu} \delta_{\nu)\alpha}\right]\\ +2\,\delta_{\alpha(\mu}p_{\nu)}+2\,\delta_{\alpha(\mu}q_{\nu)}+\delta_{\mu\nu}q_\alpha\Big\}\,.
\eesp
%From \eqref{flowtwoq}, $\Gamma^1_{2\q} = \Gamma^1_{2\q}(\mu)+I^1_{2\Lambda} - I^1_{2\mu}$. 
Choosing the complementary solution to have the same form as \eqref{B}, with kernel $\mathcal{B}^c_{\mu\nu\alpha}(p,\mu)$,  $\Gamma^1_{2\q}$ also has this form and is trivially satisfies  \eqref{mSTtwoone}. Writing its kernel as $\mathcal{B}_{\mu\nu\alpha}(p,\Lambda)$, 
%(suppressing the $\mu$ dependence) 
we have 
\be 
\label{summedB}
\mathcal{B}_{\mu\nu\alpha}(p,\Lambda) = \mathcal{B}^c_{\mu\nu\alpha}(p,\mu)+\mathcal{B}^I_{\mu\nu\alpha}(p,\Lambda)-\mathcal{B}^I_{\mu\nu\alpha}(p,\mu)\,.
\ee
The momentum integral \eqref{Bint} is a formal expression since it has quadratic and logarithmic divergences. By using dimensional regularisation to define it (using the $d$-dimensional $\cG_1$  described at the end of sec. \ref{sec:sollinear}), we automatically subtract the quadratic divergence, and by using the $\overline{\text{MS}}$ scheme we subtract the log divergence leaving just the usual $\ln\mu_R$ ambiguity.\footnote{If desired, the subtraction can be reinstated since at one loop it always appears with the same coefficient as $\ln\mu_R^2$.} 
%which we must be able to regard as part of the complementary solution. 
Taylor expanding the momentum integral up to cubic order
gives:
\besp 
\label{Bderiv}
(4\pi)^2\,I^1_{2\Lambda} = \Lambda^2\!\int^\infty_0\!\!\!\!\!\!du\, C(C-2)\,\left[\half\vp^*\partial\cu\cdot c -\tfrac98\hs_0(c^*_\mu c_\mu)\right]\\
-\half\vp^*\Box\partial\cu\cdot c
+\hs_0(\tfrac14H^*_{\mu\nu}\partial^2_{\mu\nu}\vp+\tfrac5{16}c^*_\mu\Box c_\mu)+
\frac12\int^\infty_0\!\!\!\!\!\!du\,u\,(C')^2\,\hs_0(H^*_{\mu\nu}\partial^2_{\mu\nu}\vp-\tfrac54c^*_\mu\Box c_\mu)\\
+\frac12\left(\ln\frac{\mu_R^2}{\Lambda^2}+\int^1_0\!\!\frac{du}u\,(1-C)^2+\int^\infty_1\!\!\frac{du}u\,C(C-2)\right) \hs_0(H^*_{\mu\nu}\partial^2_{\mu\nu}\vp+\tfrac34c^*_\mu\Box c_\mu)+O(\partial^5)\,,
\eesp
%where now $\mu_R$ is balanced dimensionally by $\Lambda$.
Here $C\cu=C(u)$ is the cutoff function, and we recognise amongst these expressions, instances of $\Omega_\Lambda$ \eqref{Omega} and $b$  \eqref{b}. 
The $O(\partial^5)$ and higher terms arise from UV finite integrals  (so do not depend on $\mu_R$). The derivation is sketched in app. \ref{app:derivexp}. 
As explained earlier, if we had absorbed $I_{2\mu}$ into $\Gamma_{2\q}(\mu)$, the remaining level-one part from \eqref{flowtwoq}, $\Gamma^1_{2\q}\cu=I^1_{2\Lambda}$, would already be a solution. The $\Lambda$-independent $\hs_0$-exact parts could be discarded by changing the choice of $\Gamma_{2\q}(\mu)$, but we keep them to match the  $\overline{\text{MS}}$ scheme. We only need to recognise that the end result \eqref{summedB} must be independent of $\mu_R$. Thus we set the one-loop complementary solution part to
\be 
\label{comptwoone}
\Gamma^1_{2\q}(\mu) = i\! \int_p\!\! H^*_{\mu\nu}(p)\, \mathcal{B}^c_{\mu\nu\alpha}(p,\Lambda)\, c_\alpha(-p) =  I^1_{2\mu}+ Z^1_2(\mu) \,\hs_0(H^*_{\mu\nu}\partial^2_{\mu\nu}\vp+\tfrac34c^*_\mu\Box c_\mu)\,,
\ee
which is  independent of $\mu_R/\mu$, since this dependence cancels between $I^1_{2\mu}$ and 
\be 
\label{ztwoone}
Z^1_2(\mu) = \frac{1}{(4\pi)^2}\ln\!\frac{\mu}{\mu_R}+z^1_{2}\,.
\ee
We see that $\kappa^2Z^1_2(\mu)$ induces a change of BRST cohomology representative at second order, as expected.\footnote{In general this would not be clear until we computed the $\mu_R$ dependence at all antighost levels, but see \eqref{Amu} and the discussion below it.} 
%by adding
%\be 
%\label{Bmu}
% Z^1_2(\mu) \,\hs_0(H^*_{\mu\nu}\partial^2_{\mu\nu}\vp+\tfrac34c^*_\mu\Box c_\mu) 
%\ee
%to the action. 
In this case the change is $\hs_0$-exact and thus amounts to a canonical reparametrisation \cf sec. \ref{sec:BRSTexact}, hence $Z^1_2$ is a wavefunction-like parameter. Its presence ensures that $\Gamma^1_2$ is also independent of the initial point $\mu$ on the renormalized trajectory, since a change of $\mu\mapsto \alpha\,\mu$ in the total solution $\mathcal{B}_{\mu\nu\alpha}(p,\Lambda)$ \eqref{summedB} can be absorbed by a change $\delta Z^1_2 = \delta z^1_{2} = -\ln\alpha/(4\pi)^2$. Altogether the one-loop level-one solution \eqref{flowtwoq} to the renormalized trajectory is:
\besp 
\label{BderivSelf}
(4\pi)^2\,\Gamma^1_{2\q} = \Lambda^2\!\int^\infty_0\!\!\!\!\!\!du\, C(C-2)\,\left[\half\vp^*\partial\cu\cdot c -\tfrac98\hs_0(c^*_\mu c_\mu)\right]\\
-\half\vp^*\Box\partial\cu\cdot c
+\hs_0(\tfrac14H^*_{\mu\nu}\partial^2_{\mu\nu}\vp+\tfrac5{16}c^*_\mu\Box c_\mu)+
\frac12\int^\infty_0\!\!\!\!\!\!du\,u\,(C')^2\,\hs_0(H^*_{\mu\nu}\partial^2_{\mu\nu}\vp-\tfrac54c^*_\mu\Box c_\mu)\\
+\frac12\left((4\pi)^2Z^1_2(\Lambda)+\int^1_0\!\!\frac{du}u\,(1-C)^2+\int^\infty_1\!\!\frac{du}u\,C(C-2)\right) \hs_0(H^*_{\mu\nu}\partial^2_{\mu\nu}\vp+\tfrac34c^*_\mu\Box c_\mu)+O(\partial^5)\,.
\eesp
If we work in scaled variables, where we absorb $\Lambda$ according to dimensions, the result depends on $\Lambda$ only indirectly through $Z^1_2(\Lambda)$. The scaled result is thus of self-similar form as expected for a renormalization group trajectory \cite{Morris:1998}. 
Renormalization schemes follow from the choice of renormalization condition for $Z^1_2$. For example, the $\overline{\text{MS}}$ scheme is recovered here with the renormalization condition
\be
\label{zrenormcond}
Z(\mu)=0 \quad\text{at}\quad\mu=\mu_R\,,
\ee 
which sets $z^1_{2}=0$ in \eqref{ztwoone}.
Evaluating the physical limit, $\mathcal{B}_{\mu\nu\alpha}(p) =\lim_{\Lambda\to0} \mathcal{B}_{\mu\nu\alpha}(p,\Lambda)$, a standard Feynman integral, we get for the physical vertex in the scheme \eqref{zrenormcond}
\com{997.5}
\be
\label{Bphys}
(4\pi)^2\,\mathcal{B}_{\mu\nu\alpha}(p) %= \lim_{\Lambda\to0} \mathcal{B}_{\mu\nu\alpha}(p,\Lambda) 
= %\frac1{(4\pi)^2}\left\{ 
\left(\tfrac34p^2p_\mu\delta_{\nu\alpha}-\half p_\mu p_\nu p_\alpha\right) \ln(p^2/\mu^2)+\tfrac23p_\mu p_\nu p_\alpha -\tfrac56 p^2 p_\mu \delta_{\nu\alpha}+\tfrac16\delta_{\mu\nu}p^2p_\alpha\,,
%\right\}\,.
\ee
where the net effect of the choice of complementary solution \eqref{comptwoone} and renormalization condition \eqref{zrenormcond} is just to convert $\mu_R$ to $\mu$.

At antighost level zero, the one-loop solution \eqref{flowtwoq} is now written as
\be 
\label{twopart}
\Gamma^0_{2\q} = \cG^0_{2\q2}
 + \delta\Gamma^0_{2\q}(\mu) +I^0_{2\Lambda}-I^0_{2\mu}\,,
\ee
the first two terms on the RHS being the complementary solution having split off the one-loop tadpole  \eqref{comptad}. Adopting a parallel notation to above we write  
\com{NB integral in \eqref{Aint} is defn of A in notes}
\be 
\label{A}
I^0_{2\Lambda} = \tfrac12\!\int_p\!\! H_{\mu\nu}(p)\, \mathcal{A}^I_{\mu\nu\alpha\beta}(p,\Lambda)\, H_{\alpha\beta}(-p)\,.
\ee
Here $\mathcal{A}^I_{\mu\nu\alpha\beta}(p,\Lambda)$  has two contributions: one from using two $\cG^1_1$ vertices joined by ghost propagators and one from two copies of $\cG^0_1$ joined by $H$ propagators. As a formal integral in $d\cu=4$ dimensions, and understood to be symmetrised \ie to be recast as $\mathcal{A}^I_{\left((\mu\nu)(\alpha\beta)\right)}$, we can write it as:
%over $(\alpha\beta)$ and $(\mu\nu)$ and exchange of these pairs,
\besp
\label{Aint}
\mathcal{A}^I_{\mu\nu\alpha\beta}(p,\Lambda) = \\%
\int_q\!C_\Lambda(q)\,C_\Lambda(p\cu+q)\Bigg\{
\frac{-1}{q^2(p\cu+q)^2}\Big[p_\alpha p_\beta p_\mu p_\nu
+2p_\alpha p_\beta p_\mu q_\nu 
+2p_\alpha p_\beta q_\mu q_\nu 
+p_\alpha p_\mu q_\beta q_\nu %\\
+2p_\alpha q_\beta q_\mu q_\nu
+q_\alpha q_\beta q_\mu q_\nu \\%
-p^2\delta_{\alpha\mu}p_\beta p_\nu
-\tfrac12p^2\delta_{\mu\nu}(p_\alpha p_\beta+3p_\alpha q_\beta+3q_\alpha q_\beta)
+\tfrac1{16}p^4\delta_{\mu\nu}\delta_{\alpha\beta}
+\tfrac12p^4\delta_{\alpha\mu}\delta_{\beta\nu}
\Big] %\\
+\frac{1}{q^2}\Big[ \tfrac18 p^2\delta_{\alpha\beta}\delta_{\mu\nu}
+\tfrac54p\cdot q\delta_{\alpha\beta}\delta_{\mu\nu} \\%
-p\cu\cdot\mkern-1mu(p\cu+q)\delta_{\alpha\mu}\delta_{\beta\nu}
+2\delta_{\alpha\mu} (p\cu+q)_\beta(p\cu+q)_\nu
-\delta_{\mu\nu}(p_\alpha p_\beta+3p_\alpha q_\beta+q_\alpha q_\beta)\Big] %\\
+\tfrac14\delta_{\alpha\beta}\delta_{\mu\nu}
\Bigg\}
\eesp
Again we define it however using $\overline{\text{MS}}$. Up to $O(\partial^2)$, \eqref{A} takes the form
\besp
\label{Aderiv}
(4\pi)^2\,I^0_{2\Lambda} = \Lambda^4\!\int^\infty_0\!\!\!\!\!\!du\,u\, C(C-2)\,\left[ \tfrac5{24}h^2_{\mu\nu}+\tfrac18\vp^2\right]\\ + \Lambda^2\!\int^\infty_0\!\!\!\!\!\!du\, C(C-2)\,\left[ \tfrac5{24}\vp\partial^2_{\alpha\beta}h_{\alpha\beta}+\tfrac58(\partial_\alpha h_{\alpha\beta})^2-\tfrac{19}{48}(\partial_\gamma h_{\alpha\beta})^2-\tfrac5{32}(\partial_\alpha\vp)^2\right]\\
- \Lambda^2\!\int^\infty_0\!\!\!\!\!\!du\,u^2(C')^2\,\left[ \tfrac1{12}\vp\partial^2_{\alpha\beta}h_{\alpha\beta}+\tfrac18(\partial_\alpha h_{\alpha\beta})^2+\tfrac{7}{96}(\partial_\gamma h_{\alpha\beta})^2+\tfrac1{16}(\partial_\alpha\vp)^2\right] +O(\partial^4)\,.
\eesp
It is a unique result but acquires dependence on $\ln\mu_R$, which appears amongst the $O(\partial^4)$ terms. (We do not display all these terms because there are rather too many.) Setting $\delta\Gamma^0_{2\q}(\mu)=I^0_{2\mu}$,  $\Gamma^0_{2\q}$ \eqref{twopart} would already be a solution. As before, we choose the complementary solution to be this up to converting the $\ln\mu_R$ dependence to $\ln\mu$ dependence. We find
\be 
\label{Amu}
\delta\Gamma^0_{2\q}(\mu) = I^0_{2\mu} +Z^0_{2a} (R^{(1)}_{\mu\nu\alpha\beta})^2+Z^0_{2b}(R^{(1)})^2\,,
\ee
where to one loop,
\be 
\label{Zzeros}
Z^0_{2a}(\mu) = -\frac{61}{120(4\pi)^2}\ln\!\frac{\mu}{\mu_R}+z^0_{2a}\,,\qquad
Z^0_{2b}(\mu) = -\frac{23}{120(4\pi)^2}\ln\!\frac{\mu}{\mu_R}+z^0_{2b}\,.
\ee
Again the r\^ole of these $\kappa^2Z$s is (also) to ensure that the full solution is actually independent of $\mu$ at $O(\kappa^2)$, and ensuring that the scaled result is a self-similar solution \cite{Morris:1998}. Since the only other $\ln\mu$ part, sitting in $\Gamma^1_{2\q}(\mu)$ \eqref{comptwoone}, is already $\hs_0$-closed, this addition must be $\hs_0$-closed, which it is by virtue of being invariant under linearised diffeomorphisms. 
As we saw, \eqref{exactinvariant}, it is actually $\hs_0$-cohomologically trivial, and thus as a consequence of the Koszul-Tate differential \eqref{KTHc},  vanishes on the free equations of motion (\ie on shell), making the $Z$s here also wave-function-like. This is also clear directly, on using the Gauss-Bonnet identity \eqref{GB}  \cite{tHooft:1974toh}. (Note that the coefficients do not agree with those in ref. \cite{tHooft:1974toh} which are computed in the background field method. The terms only have to agree on-shell, which they do trivially since they both vanish.) Again the $\overline{\text{MS}}$ scheme is recovered by choosing the renormalization condition \eqref{zrenormcond}. 
In the physical limit, the tadpole correction \eqref{comptad} vanishes, so once more the net effect of our renormalization condition on the choice of complementary solution \eqref{Amu} is to swap $\mu_R$ for $\mu$. We find for the physical $\Gamma^0_2$ two-point vertex
(where again we mean this to be recast as $\mathcal{A}_{\left((\mu\nu)(\alpha\beta)\right)}$):
\besp 
\label{Aphys}
(4\pi)^2\,\mathcal{A}_{\mu\nu\alpha\beta}(p) = %\frac1{(4\pi)^2}%\Bigg\{ 
\Big(
\tfrac7{10}p_\alpha p_\beta p_\mu p_\nu
-\tfrac{23}{60}p^2\delta_{\alpha\beta}p_\mu p_\nu
-\tfrac{61}{60}p^2\delta_{\alpha\mu}p_\beta p_\nu
+\tfrac{23}{120}p^4\delta_{\alpha\beta}\delta_{\mu\nu}
+\tfrac{61}{120}p^4\delta_{\alpha\mu}\delta_{\beta\nu}
\Big) \!\ln\!\left(\!\frac{p^2}{\mu^2}\!\right)\\
+\tfrac{19}{75}p_\alpha p_\beta p_\mu p_\nu
-\tfrac{1229}{1800}p^2\delta_{\alpha\beta}p_\mu p_\nu
-\tfrac{283}{1800}p^2\delta_{\alpha\mu}p_\beta p_\nu
+\tfrac{1829}{3600}p^4\delta_{\alpha\beta}\delta_{\mu\nu}
+\tfrac{283}{3600}p^4\delta_{\alpha\mu}\delta_{\beta\nu}\,,
\eesp
the quartic on the first line being the same as appears in (\ref{Amu},\ref{Zzeros}).

%dimensional regularisation as described above, and again
%the ambiguity in the subtractions parametrise the remaining freedom we need in choice of complementary solution. 
Finally, substituting $\Gamma^0_{2\q}$ \eqref{twopart} into the one-loop second-order mST  \eqref{mSTtwoq} and using the final equation in the CME relations \eqref{CMErelations} we see that\footnote{Note that the covariantisation $\cG^0_{1\q2}$ thus plays a different r\^ole from $\cG^0_{2\q2}$.} 
\be 
\label{mSTtwopart}
Q_0 \left(\Gamma^0_{2\q}-\cG^0_{1\q2}\,\right) +Q^-_0\,\Gamma^1_{2\q} =  %Q_0\left(\cG^0_{1\q2}-\cG^0_{2\q2}\right) 
-\text{Tr}\,C^\Lambda\, \cG^{(2)}_{1*}\propH \cG^{(2)}_1\, \Big|^0\,,
\ee
($\Delta\Gamma_2$ trivially vanishes) where on the RHS we retain only the antighost level zero piece. This last term has three contributions, one with $\cG^2_1$ and $\cG^1_1$ differentiated with respect to $c^*$ and (anti)ghosts, the other two using $\cG^1_1$ and its $H^*$, and either a second copy $\cG^1_1$ differentiated with respect to $H$ and $\bar{c}$, or $\cG^0_1$ where the differentials are of course both with respect to $H$. 
%both involving $\cG^1_1$ and $\cG^2_1$, in one case differentiating with respect to $c^*$ and (anti)ghosts and in the other case differentiating with respect to $H^*$ and $H$. 
The result is: 
\be 
\label{F}
 -\text{Tr}\,C^\Lambda\, \cG^{(2)}_{1*}\propH \cG^{(2)}_1\, \Big|^0 = 
  i\! \int_p\!\! H_{\mu\nu}(p)\, \mathcal{F}_{\mu\nu\alpha}(p,\Lambda)\, c_\alpha(-p)\,,
\ee
where
\besp
\label{Fint}
\mathcal{F}_{\mu\nu\alpha}(p,\Lambda) = \int_q\!C_\Lambda(q)\,C^\Lambda(p\cu+q)\Big\{\delta_{\mu\nu}p_\alpha+3\delta_{\mu\nu}q_\alpha\\ 
+\frac{1}{q^2}[2 q_\mu q_\nu (p\cu+q)_\alpha+4p_\mu p_\nu q_\alpha   
 - 2p\cu\cdot q\, (p\cu+q)_{(\mu}\delta_{\nu)\alpha} %\delta_{\alpha(\mu}q_{\nu)} 
+p\cu\cdot q\, \delta_{\mu\nu}(p\cu+q)_\alpha  
-4\delta_{\mu\nu}q_\alpha p^2 ] \Big\}.
\eesp
The above $\mathcal{A}$, $\mathcal{B}$ and $\mathcal{F}$ vertices are analogous to vertices in Yang-Mills theory, which we labelled similarly in ref. \cite{\yuji}. 
%We write
%\be
%\cG^0_{1\q2} = \tfrac12\int_p\!\! H_{\mu\nu}(p)\, \mathcal{C}_{\mu\nu\alpha\beta}(p,\Lambda)\, H_{\alpha\beta}(-p)\,,
%\ee
%in the same form as \eqref{A}, where $\mathcal{C}_{\mu\nu\alpha\beta}$ is the polynomial computable from \eqref{h}, \eqref{Gbonetwozero} and \eqref{comptad}. 
Note that $\overline{\text{MS}}$ has no effect on $\mathcal{F}_{\mu\nu\alpha}$ or the tadpole integrals, \eqref{Gbonetwozero} and \eqref{comptad}, since these are already fully regulated by the cutoff functions and thus have no $1/\epsilon$ divergences. Writing $G^{(1)}_{\mu\nu}$ \eqref{Gmunu} in momentum space as 
\be 
\label{Gp}
G^{(1)}_{\mu\nu}(p) = -G^{(1)}_{\alpha\beta\mu\nu}(p) H_{\alpha\beta}(p)\,,
\ee
we see that \eqref{mSTtwopart} is a modified Slavnov-Taylor identity for two-point vertices:
\be 
\label{STid}
\mathcal{A}_{\mu\nu\alpha\beta}\, p_\beta +
G^{(1)}_{\mu\nu\sigma\lambda}\mathcal{B}_{\sigma\lambda\alpha} = \tfrac78b\Lambda^4 (\delta_{\mu\nu}p_\alpha-2p_{(\mu}\delta_{\nu)\alpha})+\half \mathcal{F}_{\mu\nu\alpha}\,,
\ee
where the first terms on the RHS come from putting $Q_0\,\cG^0_{1\q2}$,  on the RHS and using the formula for $\cG^0_{1\q2}$ \eqref{Gbonetwozero}.
Note that in the physical limit $\Lambda\cu\to0$, the above RHS vanishes and this equation becomes the unmodified Slavnov-Taylor identity: it just says that the amplitude $\mathcal{A}$ is gauge invariant on shell, \ie up to terms proportional to the free equation of motion $G^{(1)}_{\mu\nu}\cu=0$. We have confirmed that the physical vertices, \eqref{Bphys} and \eqref{Aphys}, do indeed satisfy the physical limit of this equation. This means that if we write the IR cutoff functions in terms of the UV one, $C_\Lambda = 1-C^\Lambda$, the LHS of the above identity \eqref{STid} can be rewritten as a sum over contributions all of which are UV regulated by $C^\Lambda$ and thus well defined without further regularisation. Further manipulation similar to those in ref. \cite{\yuji} would then establish that \eqref{STid} holds exactly as an identity between the integrals (\ref{Aint},\ref{Bint},\ref{Fint}).
In fact by the Bianchi identity, $p_\mu G^{(1)}_{\mu\nu}(p)\cu=0$, it is apparent that only the last term in the physical $\mathcal{B}$ vertex \eqref{Bphys} makes a contribution. Therefore the above identity \eqref{STid} states that the part of the physical $\mathcal{A}$ vertex dependent on renormalization conditions, namely the $\ln p^2/\mu^2$ part of \eqref{Aphys}, is transverse, a property we have already established in \eqref{Amu}.
The derivative expansion of $\mathcal{F}$ \eqref{F} gives:\footnote{Again note that $\Omega_\Lambda$ (\ref{Omega}) and $b$ (\ref{b}) give alternative expressions for the terms linear in $C$.}
\besp
\label{Fderiv}
-(4\pi)^2\,\text{Tr}\,C^\Lambda\, \cG^{(2)}_{1*}\propH \cG^{(2)}_1\, \Big|^0 =  
\Lambda^4\!\int^\infty_0\!\!\!\!\!\!du\,u\, C(C-2)\,\left[ \tfrac56 h_{\mu\nu}\partial_\mu c_\nu+\tfrac14\vp\partial\cu\cdot c\right] \ -b\Lambda^4\left[\tfrac76h_{\mu\nu}\partial_\mu c_\nu+\tfrac{15}4\vp\,\partial\cu\cdot c\right]\\
+\Lambda^2\!\int^\infty_0\!\!\!\!\!\!du\, C(C-2)\,\left[ \tfrac13 h_{\mu\nu}\Box\partial_\mu c_\nu
+\tfrac{11}8\vp\Box\partial\cu\cdot c-\tfrac{11}{12}h_{\mu\nu}\partial^3_{\mu\nu\alpha}c_\alpha\right]\\
+\Lambda^2\!\int^\infty_0\!\!\!\!\!\!du\,u^2(C')^2\,\left[ \tfrac1{24}h_{\mu\nu}\partial^3_{\mu\nu\alpha}c_\alpha+\tfrac{13}{24}h_{\mu\nu}\Box\partial_\mu c_\nu\right] +O(\partial^5)\,.
\eesp
We have verified that the one-loop second-order mST identity \eqref{mSTtwopart} is satisfied up to $O(\partial^3)$ by the derivative expansions \eqref{Bderiv}, \eqref{Aderiv} and \eqref{Fderiv} together with the tadpole corrections (\ref{Gbonetwozero},\ref{comptad}).
%(\ie using (\ref{Omega},\ref{b},\ref{Gbonetwozero},\ref{comptad},\ref{Bderiv},\ref{Aderiv},\ref{Fderiv})). 
In particular this confirms explicitly that these tadpole contributions  automatically
supply required $O(\partial^{0})$ and $O(\partial^2)$ terms necessary for satisfying this identity.

\section{Discussion}
\label{sec:discussion}

We have seen that at second order in perturbation theory the end result is the standard one for the one-particle irreducible effective action at $O(\kappa^2)$, and which is thus a one loop contribution. Since we are dealing with pure quantum gravity at vanishing cosmological constant, the logarithmic running is due to wave-function-like reparametrisations. This is true in standard quantisation \cite{tHooft:1974toh} but it is also reflected in the new quantisation. 
%that in \emph{pure} gravity at second order, the logarithmic running with $\mu$ can be absorbed in reparametrisations. 
However outside the diffeomorphism invariant subspace these reparametrisations are not purely wave-function-like but are accompanied by coefficient functions, for example at antighost level zero they will  take the form:
\be 
\label{reparamln}
\delta H_{\mu\nu} = R^{(1)}_{\mu\nu} \,f^a_\Lambda(\vp,\mu)+\delta_{\mu\nu}\,R^{(1)}\,f^b_{\Lambda}(\vp,\mu)\,,\qquad \text{where}\quad  f^i_\Lambda(\vp,\mu)\to c_i\,\kappa^2\ln\mu\quad\text{as}\quad \Lambda_\p\to\infty\,,
\ee
$c_i$ being numerical constants $(i=a,b)$.
There are also infinitely many perturbative reparametrisations possible of the form 
\be 
\delta \vp = f_\Lambda(h_{\mu\nu},\vp)\,,
\ee
the RHS evidently being made up of Lorentz invariant combinations of $h_{\mu\nu}$. Some combination of these reparametrisations will correspond to redundant operators \cite{WR,Dietz:2013sba}. It is these kind of reparametrisations that would lead to a demonstration of the quantum equivalence of unimodular gravity and ordinary gravity \cite{Percacci:2017fsy,\morrii} within this new quantisation. 

Notice that the logarithmic running encapsulated in $Z^1_2(\mu)$ \eqref{ztwoone} and $Z^0_{2a,b}(\mu)$ \eqref{Zzeros}, is by no means the only logarithmic running in the theory. Infinitely many more cases are generated in the derivative expansion of the general solution for the second-order  renormalized trajectory \eqref{Gammatwosol} \cite{secondconf}. However all the other cases vanish as a power of $\Lambda_\p$ in the large amplitude suppression scale limit. 

It seems clear that once we add matter and/or a cosmological constant, it will no longer be the case that the logarithmic running inside the diffeomorphism invariant subspace
is attributable to a reparametrisation. It will have to be attributed to new diffeomorphism-invariant effective couplings. These effective couplings are precisely the same couplings that need to be introduced in standard quantisation  \cite{tHooft:1974toh}. Indeed we still expect to need a complementary solution in the form we gave for $\delta\Gamma^0_{2\q}(\mu)$ \eqref{Amu}, but the curvature-squared terms no longer vanish on the equations of motion since the Einstein tensor is now sourced by the matter stress-energy tensor and/or a term proportional to $g_{\mu\nu}$ in the case of a cosmological constant.

Actually, once inside the diffeomorphism invariant subspace, we are obeying both the flow equation \emph{and} the mST, and therefore the solution must correspond to an RG flow in the standard quantisation. The problem in standard quantisation is that these flows have an infinite number of parameters, new ones appearing at each loop order. In standard quantisation they are identified with renormalized couplings, and the corresponding bare couplings are required to absorb the UV divergences. It is clear that in this standard framework none of these flows can correspond to a genuine perturbative continuum limit in the usual Wilsonian sense, \ie a \emph{renormalized trajectory} emanating from the Gaussian fixed point, since $\kappa$ is irrelevant. (The same is true
of all higher order couplings apart from the curvature squared ones.) 

In this new quantisation we have found a solution to this latter problem: we have constructed a genuine perturbative renormalized trajectory. We have demonstrated that it works in perturbation theory, at both first order \cite{\morrii,first} and now, second order \cite{secondconf}.  It emanates from the Gaussian fixed point along relevant directions provided by the underlying (marginally) relevant couplings, $g^\sigma_{2l+\eps}$. It is these couplings that absorb the UV divergences \cite{secondconf}. Once inside the diffeomorphism invariant subspace, this {renormalized trajectory} must coincide with a subset of the RG flows derived in standard quantisation. The question is which subset. Since we need to send $\Lambda_\p\cu\to\infty$ in fig. \ref{fig:flow} to fully recover diffeomorphism invariance, we know at least that these flows must exist all the way to $\Lambda\cu\to\infty$ within the diffeomorphism invariant subspace, even though they will not qualify as part of a perturbative  {renormalized trajectory} inside this subspace.

Once inside the diffeomorphism invariant subspace, the underlying couplings disappear and the trajectory is parametrised by diffeomorphism-invariant effective couplings. One possibility is that there is no restriction: the subset is the whole set, the effective couplings are in one-to-one correspondence with the couplings required in standard quantisation. Devastating as this might be for the general predictivity of the theory, this construction suggests that there is nothing inherently inconsistent with such a scenario. 

If this is the outcome, nevertheless the new quantisation provides a different perspective. For example, it is not true that the introduction of these higher order couplings require a loss of unitarity, provided that their signs are chosen to avoid wrong-sign poles in the full propagators. In standard quantisation, the assumption is that once couplings are introduced for the curvature-squared terms for example, these couplings must be part of some `fundamental' bare action, and thus from the beginning turn the theory into one with higher derivatives even at the free (bilinear) level. Here, the bare action lies outside the diffeomorphism invariant subspace. The higher derivative interactions there must always be accompanied by a $\dd\Lambda{n}$ operator,  and thus cannot alter the kinetic terms. In other words, the bilinear action maintains its two-derivative form \cite{\morrii}. 

It remains the case that ultimately the perturbative development of the theory is organised in powers of $\kappa$ and therefore by dimensions, accompanied by increasing numbers of space-time derivatives at higher order.  But since we are dealing with a theory with a genuine continuum limit, the fact that perturbation theory breaks down in the regime\footnote{Here $\partial$ stands for the typical magnitude of space-time derivatives.} $\kappa\partial>1$, just indicates that the theory becomes non-perturbative in this regime and not, as usually interpreted, a signal of breakdown of an effective quantum field theory description. 

We see very clearly that it is the logarithmically running terms and their finite part ambiguities,  necessarily BRST invariant, that demand the introduction of new couplings 
%(over and above $\kappa$ and the cosmological constant) 
order by order in perturbative quantum gravity. In contrast, the power-law $\Lambda$ dependence is computed unambiguously. Nothing within perturbation theory demands that new couplings be associated to such $\Lambda^{2n}$ terms (integer $n\cu>0$). Nor is the field dependence associated to $\Lambda^{2n}$, closed under BRST, but rather is intimately related to the modifications of the Slavnov-Taylor identities. Thus the problem in quantum gravity is to find the \emph{mechanism}, if there is one, that determines (some or all of) the finite parts associated to the $\ln(\Lambda/\mu)$ terms that appear at the perturbative level. If for example, all these parameters are fixed by such a mechanism, we would be left with only one new parameter  at the quantum level, the mass scale that arises by dimensional transmutation from the very existence of the RG (the equivalent to $\Lambda_{QCD}$ in QCD). 

In fact we know that at third order, 
the first-order couplings will run with $\Lambda$  \cite{secondconf}. It is conceivable that this running and the required subsequent matching into the diffeomorphism invariant subspace, plays a r\^ole in providing this missing mechanism. Below, we discuss another possibility, some hints that this mechanism arises solely from insisting that the RG flow within the diffeomorphism invariant subspace, remains non-singular all the way to $\Lambda\cu\to\infty$. 
%However these constraints may  only be visible at the non-perturbative level. 
One such well-studied possibility is a non-perturbative (asymptotically safe) UV fixed point \cite{Weinberg:1980,Reuter:1996,Bonanno:2020bil}. However note that our current construction was born from attempts to solve issues with the degeneration of the fixed points and eigenoperator spectrum that are seen in that scenario if one goes (sufficiently carefully) beyond truncations involving just a finite number of operators  (see the final discussions in refs. \cite{Dietz:2016gzg,first}). As we now explore, a mechanism for fixing the parameters
could follow from the same mathematical properties of the partial differential flow equations that lead to these problems in the first place.

\section{A possible non-perturbative mechanism}
\label{sec:missing}

In the conformal sector the infinite number of couplings $g^\sigma_{2l+\eps}$ lead to a new effect, namely the fact that almost always, even at the linearised level, RG flows towards the IR become singular and then cease to exist \cite{\morri}. This is very much interwoven into the subsequent development \cite{\morrii,first,secondconf}. Indeed it is for this reason that the construction requires the initial point $\mu$ for the renormalized trajectory \eqref{Gammatwosol} to lie below $\Lambda_\p$, most of the trajectory then being safely developed from the IR to the UV. This is due to the fact that we are dealing with solutions of a parabolic partial differential equation that are non-polynomial in the amplitude: such solutions are only guaranteed when flowing from the IR to the UV \cite{\morri}. 

These comments  apply equally well to the $h_{\mu\nu}$ sector however with the crucial difference that there the equation is reverse parabolic, with solutions only guaranteed when flowing from the UV to the IR \cite{\morri}. The problem is not seen for polynomial linearised solutions, because such solutions are a finite sum of eigenoperators (the Hermite polynomials) \cite{\morri,\morrii} with constant coefficients. But diffeomorphism invariance, which is imposed in the IR (inside the diffeomorphism invariant subspace), requires us to use solutions that are non-polynomial in the $h_{\mu\nu}$ amplitude (because the curvature terms require both the metric $g_{\mu\nu}$ and the inverse metric $g^{\mu\nu}$). Thus diffeomorphism invariance forces us to consider  solutions non-polynomial in $h_{\mu\nu}$, evolving from the IR to the UV. Such solutions almost always fail at some critical scale $\Lambda_\text{cr}$, before we reach $\Lambda\cu\to\infty$.

In reality, the solution must exist simultaneously in both the $h_{\mu\nu}$ and $\vp$ sectors. Consider a solution $\delta\Gamma$ to the linearised flow equation \eqref{flowone}. Isolating the $h_{\mu\nu}$ and $\vp$ amplitude dependence, we can expand $\delta\Gamma$ over monomials $\varsigma_{\mu_1\cdots\mu_n}$:
\be 
\label{dGsol}
\delta\Gamma = \sum_{\varsigma} \varsigma_{\mu_1\cdots\mu_n}(\partial,\partial\vp,\partial h,c,\Phi^*)\, f^\varsigma_{\Lambda\mu_1\cdots\mu_n}(h_{\alpha\beta},\vp)+\cdots\,,
\ee
where we suppress Lorentz indices on the arguments in $\varsigma$ and we mean that its (anti)field arguments can appear as indicated or differentiated any number of times. These new coefficient functions $f^\varsigma_\Lambda$ are necessarily non-polynomial in $h_{\alpha\beta}$ and $\vp$ for the reasons we have explained. The linearised flow equation \eqref{flowone} can be solved exactly using the same integrating factor as in the general solution \eqref{complementary}. The ellipses in \eqref{dGsol} refer to the tadpole corrections so formed by attaching propagators to $\varsigma$ either exclusively, or also to $h_{\alpha\beta}$ and $\vp$. Now from the linearised flow equation  \eqref{flowone}, the coefficient functions themselves satisfy the flow equation:
\be 
\label{flowtotal}
\dot{f}^\varsigma_{\Lambda\mu_1\cdots\mu_n}(h_{\alpha\beta},\vp)\ =\ \Omega_\Lambda \left( \frac{\partial^2}{\partial h^2_{\mu\nu}}-\frac{\partial^2}{\partial\vp^2}\right) f^\varsigma_{\Lambda\mu_1\cdots\mu_n}\,.
\ee
Here we clearly see the property that the equation in each sector separately is parabolic, but in opposite directions, and thus in fact the Cauchy initial value problem\footnote{\TRM{This is a property of the flow under effective cutoff $\Lambda$. It has nothing to do with the existence (or otherwise in some approaches \cite{Ambjorn:2015qja,Chaney:2015mfa,Steinacker:2017vqw,Stern:2018wud,Perry:1993ry,Bojowald:2016itl,Bojowald:2018xxu}) of a Cauchy initial value surface in the dynamics of the theory. }} for such a partial differential equation is not well defined in either direction. This mathematical property is not cured, but only obscured, by using the full non-linear flow equations. We see that we are dealing with novel partial differential equations whose solution typically becomes singular when it is evolved in either direction, \emph{even at the linearised level}. %This property could be the missing mechanism that fixes the finite part ambiguities $\delta\Gamma^{(\ell)}$ that appear at $\ell$-loop order. 
As we have emphasised already for flows towards the IR in the $\vp$ sector \cite{\morri}, this does not mean solutions do not exist but rather that the initial conditions must be very special, \ie lie within a heavily restricted subspace. Below we uncover hints that this allows only the cosmological constant and $\kappa$ ultimately to exist as independent couplings.

Notice that this issue applies only to the fields that are differentiated in the flow equation, \ie to the quantum fields -- whose second order differentials together with the RG time derivative make the equations (reverse) parabolic.  It does not apply to the antifields, nor to background fields if the background field approach is followed. In fact it does not apply to the ghost fields either because these are Grassmann and thus dependence on their amplitude is necessarily polynomial. Therefore the issue only arises for the quantum fluctuation fields $h_{\mu\nu}$ and $\vp$.

To take these arguments a little further, we recall that the finite part ambiguity $\delta\Gamma_{\!\!(\ell)}$ that appears at $\ell$-loop order, is a local $\Lambda$-independent operator, and note that its dimension is 
\be 
\label{dimdG}
[\delta\Gamma_{\!\!(\ell)}]=2(\ell\cu+1)
\ee 
(\eg as required by dimensions from the factors of $\kappa$). We also note that if the mST \eqref{mST} is to be obeyed inside the diffeomorphism invariant subspace, we must have $(\Gamma_0,\delta\Gamma_{\!\!(\ell)})=0$  (since all the other parts are at higher loop order, in particular the correction term in the mST carries an extra loop) \cite{\yuji}. In other words, at $\ell$-loop order the ambiguous parts  $\delta\Gamma_{\!\!(\ell)}$ must be invariant under the full classical BRST transformations \cite{\yuji}, \cf sec. \ref{sec:solCME}, reflecting standard treatments \cite{ZinnJustin:1974mc,ZinnJustin:1975wb,ZinnJustin:2002ru}. In particular the level zero part, $\delta\Gamma^{0}$, must be diffeomorphism invariant,  and thus at one loop are curvature-squared terms, as confirmed in $\delta\Gamma^0_{2\q}$ \eqref{Amu}, at two-loop order are $\kappa^2$ times curvature cubed, or $\kappa^2R\nabla^2R$ type terms, and so forth. They are therefore indeed non-polynomial in $h_{\mu\nu}$ (and also $\vp$ as also imposed by the new quantisation).

At loop-order higher than $\ell$, where $\delta\Gamma_{\!\!(\ell)}$ first appears, $\delta\Gamma_{\!\!(\ell)}$ gets altered by the flow equation \eqref{flow} and mST \eqref{mST} in ways that are not straightforward to analyse. If we model the situation by just taking the linearised flow equation \eqref{flowone} and imposing $\delta\Gamma=\delta\Gamma_{\!\!(\ell)}$ at $\Lambda\cu=0$, the perturbation will no longer satisfy BRST invariance or the mST once $\Lambda\cu>0$. However we will be able to see the restrictions that arise from the fact that the flows are typically singular. In close similarity to the solution for the pure-$\vp$ coefficient functions \eqref{coeffgen}, the partial differential equation \eqref{flowtotal} is solved formally by the Fourier transform:
\be 
\label{fouriertotal}
f^\varsigma_{\Lambda\mu_1\cdots\mu_n}(h_{\alpha\beta},\vp) = 
 \int\frac{d^9\vpi_{\alpha\beta}\,d\vpi}{(2\pi)^{10}}\ \ff^{\varsigma}_{\mu_1\cdots\mu_n}\!(\vpi_{\alpha\beta},\vpi)\ {\rm e}^{\frac{1}{2}\Omega_\Lambda(\vpi^2_{\mu\nu}-\vpi^2)
 +i\vpi_{\mu\nu}h_{\mu\nu}+i\vpi\vp}\,,
\ee
where $\vpi_{\mu\nu}$ is  traceless, being the momentum conjugate to $h_{\mu\nu}$.  That \eqref{fouriertotal} is the Fourier form of the solution, can be seen straightforwardly by substitution, and matches the general linearised functional solution \eqref{complementary} as one can see by substituting the $\Lambda\cu=0$ Fourier transform for the physical coefficient function. However for the above to be more than a formal solution to \eqref{flowtotal}, we need the Fourier integral to converge. We see that as $\Lambda$ increases from zero, convergence in the $\vp$ sector only improves, since it is weighted by ${\rm e}^{-\frac{\vpi^2}2\Omega_\Lambda}$, reflecting the fact that the Cauchy initial value problem is well defined in this sector for IR$\to$UV  \cite{\morri}. However in the $h_{\mu\nu}$ sector the integral has the exponentially growing weight, ${\rm e}^{\frac{\vpi^2_{\mu\nu}}2\Omega_\Lambda}$. Unless $\ff^\varsigma$ decays faster than an exponential of $\vpi^2_{\mu\nu}$ (at fixed $\vpi$), the solution \eqref{fouriertotal} will be singular at some critical scale $\Lambda\cu=\Lambda_\text{cr}\cu\ge0$, above which the flow ceases to exist.

We see therefore that the flows will exist only for carefully chosen parametrisations of the metric in terms of $h_{\mu\nu}$ and $\vp$. Now we show that solutions of the form \eqref{fouriertotal} cannot exist simultaneously for all the $\delta\Gamma$ that match diffeomorphism invariant $\delta\Gamma_{\!\!(\ell)}$ at $\Lambda\cu=0$. If we take the Einstein-Hilbert action \eqref{geom} as an example and expand it over monomials as in \eqref{dGsol}, the required strong suppression of high conjugate momenta $\vpi_{\mu\nu}$ in  $\ff^\varsigma$, means that for the above to be a solution, there must be no rapid variation of the Einstein-Hilbert action under changes in the $h_{\mu\nu}$ amplitude. Obviously, at a minimum we then need a parametrisation that exists for all amplitudes. That is not true of the simple linear split form of $g_{\mu\nu}$ \eqref{gH} which is not positive definite for all $h_{\mu\nu}$ and $\vp$, and for which $g_{\mu\nu}$ is singular at $\kappa\vp=-2$, and whenever $\kappa h_{\mu\nu}$ has $-1$ as an eigenvalue. We can cure this by for example parametrising the metric $g_{\mu\nu}$ in terms of an exponential of $\kappa h_\mu{}^\nu$ (considered as a matrix), see \eg \cite{Kawai:1993mb,Eichhorn:2013xr,Nink:2014yya,Percacci:2015wwa,Percacci:2016arh}. Such a parametrisation can also ensure that the square root, in the measure $\sqrt{g}$, does not lead to branch cuts (as also  would expressing the metric in terms of a vierbein, since the measure is then its determinant).

This is still not enough to allow a solution in the form \eqref{fouriertotal} however. From the already required faster than exponential decay, we see that the mod-squared amplitudes $|\ff^{\varsigma}_{\mu_1\cdots\mu_n}|^2$ are integrable. Thus by Parseval's theorem, the squared coefficient functions $(f^\varsigma_{\Lambda\mu_1\cdots\mu_n})^2$ must also be integrable over $d^9h_{\alpha\beta}d\vp$. This in turn implies that the coefficient functions $f^\varsigma_{\Lambda\mu_1\cdots\mu_n}$ must vanish as $h_{\alpha\beta}\cu\to\infty$.\footnote{They must decay faster than $1/|h_{\alpha\beta}|^{9/2}$. Given an appropriate choice of $\ff^\varsigma$, one can get a much improved estimate by using the method of  steepest descents in \eqref{fouriertotal}.} Since $\sqrt{g}R\cu\mapsto \alpha \sqrt{g}R$ under scaling $g_{\mu\nu}\cu\mapsto \alpha g_{\mu\nu}$ (where $\alpha$ is some constant), we see that this last condition will hold true for the Einstein-Hilbert action if and only if $g_{\mu\nu}$ itself vanishes in this limit. 

A Fourier solution \eqref{fouriertotal} for the cosmological constant term, is then not ruled out by this condition, since $\sqrt{g}\mapsto\alpha^2\sqrt{g}$, and thus it will also vanish in the limit $h_{\alpha\beta}\cu\to\infty$. However all the higher derivative terms are then ruled out from having such solutions, since curvature-squared terms go like $\alpha^0$, while the higher order terms behave as negative powers of $\alpha$ and thus actually diverge in the limit $h_{\alpha\beta}\cu\to\infty$.

Notice that despite the fact that we are modelling using only linearised solutions, the arguments we are making are non-perturbative in $\kappa$, because the breakdown in the solutions happens at finite or diverging $\kappa h_{\mu\nu}$. In general the level-zero part satisfies $\delta\Gamma_{\!\!(\ell)}  \cu\mapsto \alpha^{1-\ell} \delta\Gamma_{\!\!(\ell)}$, and thus if these perturbations had to extend to solutions $\delta\Gamma$ of Fourier type \eqref{fouriertotal}, we would have shown that, despite the apparent freedom to change individually the new effective couplings that appear at each loop order, non-perturbatively in $\kappa$ the requirement that the renormalized trajectory is non-singular actually rules out all such infinitesimal changes $\delta\Gamma_{\!\!(\ell)}$. We would therefore conclude that the only freely variable couplings are in fact $\kappa$ itself and the cosmological constant. 

We cannot quite draw such dramatic conclusions however. The arguments we have presented can only be regarded as hints. 
Firstly, solutions exist to the linearised flow equations \eqref{flowtotal} that do not fit the assumed Fourier form \eqref{fouriertotal}. For example solutions polynomial in the graviton can be cast in Fourier space, but $\ff^\varsigma$ is then distributional, \viz a sum over differentials of $\delta(\vpi_{\alpha\beta})$. Another example is provided by the $\vp$ part of exponential parametrisation \cite{Kawai:1993mb,Eichhorn:2013xr,Nink:2014yya,Percacci:2015wwa,Percacci:2016arh} which extends to the solution
\be 
\label{expphiflow}
f(\vp) =  {\rm e}^{\frac{\kappa}{2}\vp} \ \implies\ 
f_\Lambda(\vp) 
%=\int^\infty_{-\infty}\!\!\!\!\!\! d\ph_0\  {\rm e}^{\frac{\kappa}{2}\vp}\, \ddp\Lambda0{\vp\cu-\vp_0} \ 
= {\rm e}^{\frac{\kappa}{2}\vp+\frac18\kappa^2\Omega_\Lambda}\,,
\ee
as can be confirmed by direct substitution in \eqref{flowtotal} or by using the Green's function   $\ddp\Lambda0{\vp\cu-\vp_0}$, \cf \eqref{physical-dnL} \cite{\morri,first}. However these are not sufficient to parametrise the Einstein-Hilbert action. In fact finding a parametrisation that can be extended to a solution of the linearised flow equation \eqref{flowtotal}, either of Fourier type \eqref{fouriertotal} or otherwise, looks challenging.\footnote{We did not find a  parametrisation of $g_{\mu\nu}$ that leads to $\ff^\varsigma$ with decay faster than exponential of $\vpi^2_{\mu\nu}$. Approaching from the other direction, nor did we find such $\ff^\varsigma$ that then lead to a non-singular $g_{\mu\nu}$.}
It is even more challenging to find one that also works for the cosmological constant term, and it is not credible that a parametrisation could be found that would also allow solutions for the higher derivative terms $\delta\Gamma_{\!\!(\ell)}$. On the contrary, it may be that there is no sensible solution even for the Einstein-Hilbert action alone. Secondly, infinitesimal changes $\delta\Gamma_{\!\!(\ell)}$ do not in fact have to satisfy the simple linearised equations \eqref{flowtotal} but operator flow equations that depend on the rest of the effective action:
\be 
\label{flowOp}
\delta\dot{\Gamma}_{\!\!(\ell)} = \half\, \text{Str}\left( \dot{\prop}_\Lambda\propH^{-1} \left[1+\propH \Gamma^{(2)}_I \right]^{-1}\propH \delta\Gamma_{\!\!(\ell)}^{(2)} \left[1+\propH \Gamma^{(2)}_I \right]^{-1}\right) 
\,,
\ee
as follows immediately from perturbing the exact RG flow equation \eqref{flow}. However, although these flow equations are much more involved than the simple linearised flow equations \eqref{flowtotal}, and are such that they allow solutions that remain compatible with BRST invariance through the (perturbed) mST \eqref{mST}, they share with \eqref{flowtotal} the property that their Cauchy initial value problem is not well defined in either direction.

\section{\TRM{General gauges}}
\label{sec:gauge}

\TRM{
All the results in this paper were derived in Feynman -- De Donder gauge. 
%it is legitimate to wonder how the framework changes in another gauge, in particular whether any part of the structure is actually dependent on this special gauge choice. 
In this section we will show that the structure changes only in an inessential way for a class of  gauge conditions.
First we recall that a great advantage of using off-shell BRST invariance is that BRST invariant correlators are independent of the choice of gauge \cite{Batalin:1981jr,Batalin:1984jr,\morrii}. Indeed in terms of the quantum fields, $\phi^A$, and the Wilsonian effective action, $S$, the mST \eqref{mST} is just the Quantum Master Equation (QME)
\be 
\label{QME}
\half (S,S) -\Delta S=0\,,
\ee
whose powerful algebraic properties continue to hold exactly despite regularisation by the cutoff function \cite{\morrii}. In more detail, an operator $\Op_i$ is (off-shell) BRST invariant if it satisfies
\be 
\label{fullBRS}
s\,\Op_i = (S,\Op_i) -\Delta\Op_i = 0\,,
\ee
while a change of gauge is implemented by adding a BRST exact term $s\, \delta\! K$ to the action \cite{Batalin:1981jr,Batalin:1984jr}. Then it follows that a BRST invariant correlator is invariant under change of gauge \cite{Batalin:1981jr,Batalin:1984jr,\morrii}:
\be 
\label{vanishingCorrelator}
\delta \langle  \Op_1\cdots\Op_n \rangle = 
- \langle s\delta\! K\, \Op_1\cdots\Op_n \rangle =
- \langle s \left(\delta\!  K \Op_1\cdots\Op_n \right) \rangle = \frac1{\mathcal{Z}}
\int\!\!\mathcal{D}\phi\, \Delta \left( \delta\! K \Op_1\cdots\Op_n\,  {\rm e}^{-S} \right) =0\,.
\ee
($\mathcal{Z}$ is the normalisation of the partition function. In this last step we use algebraic properties of the QME, the disjoint support of the $\Op_i$, and the fact that $\Delta$ now contains a total functional derivative with respect to $\phi$ \cite{\morrii}.) 

The formulation we are using here is entirely equivalent since the two formulations are mapped into each other by the Legendre transform relation \cite{Morris:1993,Ellwanger1994a,Rosten:2010pc,Morris:2015oca,\yuji}:
%,Bonini:1992vh,Keller:1990ej}:
\be 
\label{Legendre}
\Gamma_I[\Phi,\Phi^*] = S_I[\phi,\Phi^*] - \half\, (\Phi -\phi)^A\, \prop^{-1}_{\Lambda\, AB}\,(\Phi-\phi)^B\,,
\ee
where $S_I$ is the corresponding interaction part of the Wilsonian effective action. Even so, since the $\Phi^A$ are not BRST invariant operators, $\Gamma$ will now depend on the choice of gauge in some unilluminating way.\footnote{\TRM{If instead we used this to compute the Schwinger functional of only gauge invariant operators we would still find results that are gauge parameter independent.}}
However in the physical limit $\Lambda\to0$, the mST \eqref{mST} becomes  the Zinn-Justin equation and is obeyed exactly. Now we can also go on shell and at this point the $\Phi^A$ do provide BRST invariant states. Thus the on-shell vertices of the physical $\Gamma$ are independent of the choice of gauge. These vertices obey the unmodified Slavnov Taylor identities, as we have already seen from \eqref{STid}. Indeed we saw that this just tells us that the physical amplitudes are gauge invariant on shell.

While gratifying, these results are hardly unexpected. After all we saw in sec. \ref{sec:evaluation}  that the second order equations \eqref{flowtwocl} -- \eqref{mSTtwocl} are identical to those we would derive in standard quantisation at one loop in this framework. All this follows provided the flow finishes up inside the diffeomorphism invariant subspace, \ie
provided the QME \eqref{QME}, equivalently mST \eqref{mST}, holds exactly in the infrared. However a central feature of our construction is that the renormalized trajectory lies outside this subspace for $\Lambda>\Lp$. Thus the important question is what happens to this part of the trajectory in other gauges, in particular whether it continues to be supported by a novel tower of relevant operators in the ultraviolet, and whether the trajectory still enters the diffeomorphism invariant subspace in the infrared.

To investigate this we rederive the flow of the upper part of the trajectory in a more general De Donder gauge. This  gauge is implemented by using the gauge fixing functional
\be 
\label{DeDonder}
F_\mu = \partial_\nu H_{\nu\mu} -\partial_\mu\vp\,.
\ee
Although we use the Batalin-Vilkovisky framework \cite{Batalin:1981jr,Batalin:1984jr}  to implement off-shell BRST invariance, for the graviton sector the general De Donder gauge amounts to adding to the free graviton action $\Gamma_0$, \cf \eqref{Gzero}, the term $\half\alpha F^2_\mu$, where $\alpha$ is the gauge parameter. Up until now we have used Feynman -- De Donder gauge, $\alpha=2$, since such a choice leads to significant simplifications:
\be 
\label{FDD}
\Gamma_0|_\text{Feynman De Donder} \equiv \half \left(\partial_\lambda h_{\mu\nu}\right)^2 -\half \left(\partial_\lambda \ph\right)^2\,,
\ee
in particular decoupling the conformal mode $\vp$ from the traceless part $h_{\mu\nu}$. Our
construction is built on a succession of results reported in previous papers \cite{\morri,Kellett:2018loq,Morris:2018upm,\morrii,first,secondconf}, where also the  Feynman -- De Donder gauge was used. Therefore we need to go back to the beginning to show how things now change in a more general gauge.

% where in this last step we use identities obeyed by quantum measure operator $\Delta$

%interaction parts of each effective action are related via the Legendre transform relation \cite{Morris:1993,Ellwanger1994a,Rosten:2010pc,Morris:2015oca}:
%%,Bonini:1992vh,Keller:1990ej}:
%\be 
%\label{Legendre}
%\Gamma_I[\ph,\ph^*] = S_I[\Ph,\Ph^*] - \half\, (\Ph -\ph)^A\, \propH^{-1}_{ AB}\,(\Ph-\ph)^B\,.
%\ee

%we recall that the formalism wri

%Using the Batalin-Vilkovisky formalism and off-shell BRST invariance involves some  complications because we need antifields, add further fields to extend to a non-minimal gauge invariant basis by adding the antighost antifield $\bar{c}^*_\mu$ and an auxiliary field $b_\mu$ which mixes with $H_{\mu\nu}$, and then gauge fix. However to all orders in perturbation theory,  no interactions are generated between these new fields, while the antighost $\bar{c}_\mu$ only appears in the combination given on the right hand side of \eqref{gaugeFixed}. 

% is as we saw in \cite{first,\yuji},
%
%to introduce instead an auxiliary field 
%
% which however we have seen can be largely avoided
%
%The construction in this paper was built on a succession of results reported in previous papers \cite{\morri,Kellett:2018loq,Morris:2018upm,\morrii,first,secondconf}, where also the  Feynman - De Donder gauge was used. Therefore we need to go back to the beginning to show how the construction changes in a more general gauge. 

The central observation that led to the new quantisation is that in Euclidean signature, the Einstein-Hilbert action is unbounded from below. This is a gauge invariant statement: the unboundedness is caused by the fact that the action is proportional to the scalar curvature, $R$, which can take either sign of any magnitude.\footnote{\TRM{It is actually positive scalar curvature that is  the problem \cite{Gibbons:1978ac,\morri}.}} 
At the free level $\Gamma_0$ still has these problems. In the Feynman De Donder gauge \eqref{FDD} this is particularly clear. Before gauge fixing the situation is obscured by linearised gauge invariance $\delta H_{\mu\nu} = \partial_\mu \xi_\nu+\partial_\nu\xi_\mu$, equivalently linearised BRST \eqref{QH}. However the gauge invariant statement is that the action \eqref{Gzero} is unbounded below in the following direction \notes{690.4,690-2}
\be 
2\vp - \frac{\partial^2_{\alpha\beta}}{\Box}\,H_{\alpha\beta} = 
2\left(1-\frac1d\right)\vp - \frac{\partial^2_{\alpha\beta}}{\Box}\, h_{\alpha\beta} = \frac1{-\Box} R^{(1)}\,,
\ee
where we used \eqref{curvatures}.
Using different gauge choices we can shift the instability to different modes, but we cannot remove it. Indeed in the Landau gauge limit of the De Donder gauge \eqref{DeDonder}, where we insist that $F_\mu = 0$ identically, the conformal mode coincides with this (linearised) gauge invariant quantity:
\be 
\vp =  \frac1{-\Box} R^{(1)}\,.
\ee
By adapting the off-shell BRST Batalin-Vilkovisky framework, what we actually do is first 
go to the non-minimal gauge invariant basis by adding to the action
\be 
\label{nonminimal}
\frac1{2\alpha}b^2_\mu-ib_\mu \bar{c}^*_\mu\,,
\ee
where  $\bar{c}^*_\mu$ is the antighost antifield and  $b_\mu$ is the auxiliary field \cite{\morrii,\yuji,first}. Adding a BRST exact term involving the gauge fixing fermion \cite{Batalin:1981jr,Batalin:1984jr} $\Psi = \bar{c}_\mu F_\mu$, then induces a canonical transformation to the gauge fixed basis. The map is the one we already gave in equation \eqref{gaugeFixed}, together with \cite{\morrii}
\be 
\label{cstargf}
\bar{c}^*_\mu \,|_\text{gi}  = \bar{c}^*_\mu \,|_\text{gf} - F_\mu\,.
\ee
Since $\Psi$ does not involve $\alpha$, these are not affected by the more general gauge. However the shifts allow the kinetic terms to be inverted to give the propagators, now in general gauge $\alpha$:
\beal
\langle b_\mu(p) \,H_{\alpha\beta}(-p)\rangle &= -\langle H_{\alpha\beta}(p)\,b_\mu(-p)\rangle  = 2\, \delta_{\mu (\alpha} p_{\beta)}/{p^2}\,,\\
\langle b_\mu(p)\,b_\nu(-p)\rangle &= 0\,,\\
\label{hpalph}
\prop_{\mu\nu\ph}(p):=\langle h_{\mu\nu}(p)\, \ph(-p)\rangle &= \langle \ph(p)\,h_{\mu\nu}(-p)\rangle =
\left(1-\frac2\alpha\right) \left( \frac{\delta_{\mu\nu}}{d}-\frac{p_\mu p_\nu}{p^2}\right) \frac1{p^2}\,,\\
\label{ppalph}
\prop_{\ph\ph}(p):=\langle \ph(p)\,\ph(-p)\rangle &= \left(\frac1\alpha - \frac{d-1}{d-2}\right) \frac1{p^2}\,,
\\
\prop_{\mu\nu\,\alpha\beta}(p):=\langle h_{\mu\nu}(p)\,h_{\alpha\beta}(-p)\rangle &= \frac{\delta_{\mu(\alpha}\delta_{\beta)\nu}}{p^2} 
+\left(\frac4\alpha-2\right)\frac{p_{(\mu}\delta_{\nu)(\alpha}p_{\beta)}}{p^4}
+\frac1{d^2} \left(\frac4\alpha-d-2\right) \frac{\delta_{\mu\nu}\delta_{\alpha\beta}}{p^2}\nn\\
\label{hhalph}
&\qquad\qquad\qquad\qquad+\frac2d\left(1-\frac2\alpha\right)\frac{\delta_{\alpha\beta}p_\mu p_\nu+p_\alpha p_\beta \delta_{\mu\nu}}{p^4}\,.
\eeal
These generalise the Feynman gauge results \eqref{HH}--\eqref{pp}. Comparing the new $h_{\mu\nu}$ propagator \eqref{hhalph} to the old one \eqref{hh}, underlines why it is preferable to work in Feynman gauge. Note that $b_\mu$ does not actually propagate into itself. The $\half\alpha F^2_\mu$ term mentioned earlier would be generated by integrating out $b_\mu$ after the transformation \eqref{cstargf}. The ghost propagator is not displayed since it is unchanged from \eqref{cc}.

Now we recall that to all orders in perturbation theory,  no interactions are generated involving the  extended basis,  $b_\mu$ and $\bar{c}^*_\mu$,   while the antighost $\bar{c}_\mu$ only appears in the combination given on the right hand side of \eqref{gaugeFixed} \cite{\yuji,first}. These statements follow from the fact that the first order interaction $\Gamma_1$ can be constructed from the minimal set (see the final paragraphs of sec. 2 in ref. \cite{first}). We will confirm that this still holds shortly.
This means that we can continue to work in minimal gauge invariant basis, applying \eqref{gaugeFixed} only while computing the ghost propagator corrections. 

The first step is to solve the linearised flow equation \eqref{flowone} to find the eigenoperators, now in general De Donder gauge. Since $h_{\mu\nu}$ and $\ph$ now propagate into each other, we need an expansion over monomials with coefficient functions involving both $h_{\mu\nu}$ and $\ph$. In other words we have an expansion
which is actually identical to that considered in \eqref{dGsol}:
\be 
\Gamma_1 = \sum_{\varsigma} \varsigma_{\mu_1\cdots\mu_n}(\partial,\partial\vp,\partial h,c,\Phi^*)\, f^\varsigma_{\Lambda\mu_1\cdots\mu_n}(h_{\alpha\beta},\vp)+\cdots\,.
\ee
Again the ellipses refer to tadpole corrections formed by attaching propagators to $\varsigma$ either exclusively, or also to $h_{\alpha\beta}$ and $\vp$. Once again, the linearised flow equation is solved exactly using the same integrating factor as in the general solution \eqref{complementary}:
\be 
\label{compGen}
\Gamma_1 = 
\exp\left(-\frac12 {\prop}^{\Lambda\,AB} \frac{\partial^2_l}{\partial\Phi^B\partial\Phi^A}\right)\,
\Gamma_{\text{phys}}\,,
\ee
where $\Gamma_{\text{phys}}$ is the $\Lambda\to0$ limit \eqref{physical}. From \eqref{flowone}, the coefficient functions satisfy the flow equation 
\be 
\dot{f}^\varsigma_{\Lambda\mu_1\cdots\mu_n}(h_{\alpha\beta},\vp)\ =\ \int_p\left( 
\prop^\Lambda_{\mu\nu\,\alpha\beta}(p) \frac{\partial^2}{\partial h_{\mu\nu}\partial h_{\alpha\beta}}
+2\prop^\Lambda_{\mu\nu \ph}(p) \frac{\partial^2}{\partial h_{\mu\nu}\partial\ph}+\prop^\Lambda_{\ph\ph}(p)\frac{\partial^2}{\partial\vp^2}\right) f^\varsigma_{\Lambda\mu_1\cdots\mu_n}\,,
\ee
where we set $d=4$ and used the fact that the tadpole integrals have quadratic dependence on $\Lambda$.
Performing the tadpole integrals we see that the $\ph$ $h_{\mu\nu}$ cross-term vanishes. (It has to because there is no invariant traceless rank two tensor.) Computing the other two we find
\be 
\dot{f}^\varsigma_{\Lambda\mu_1\cdots\mu_n}(h_{\alpha\beta},\vp)\ =\  \Omega_\Lambda \left[ \left(\frac1\alpha+\frac12\right) \frac{\partial^2}{\partial h^2_{\mu\nu}}+\left(\frac1\alpha-\frac32\right)\frac{\partial^2}{\partial\vp^2}\right] f^\varsigma_{\Lambda\mu_1\cdots\mu_n}\,,
\ee
where $\Omega_\Lambda$ was defined in \eqref{Omega}. Thus the flow of the coefficient function is again parabolic of a simple form in each sector separately. If we choose 
\be 
\label{validalpha}
\alpha>\frac23\qquad\text{or}\qquad \alpha<-2\,, 
\ee
the sign on the right hand side is negative for $\ph$ and positive for $h_{\mu\nu}$ just as before. Therefore we find \cite{\morrii} that the eigenoperators we have to expand in take the same form \eqref{topFull} as before. In particular the sum over eigenoperators converges (in the square integrable sense) for otherwise arbitrary coefficient functions, only if the interactions are expanded over polynomials in $h_{\mu\nu}$ times the operators $\dd{\Lambda}{n}$. The only difference is that these latter operators now involve a rescaled $\Omega_\Lambda$:
\be
\label{flowalpha}
\dd{\Lambda}{n} := \frac{\partial^n}{\partial\vp^n}\, \dd{\Lambda}{0}\,, \qquad{\rm where}\qquad \dd{\Lambda}0 := \frac{1}{\sqrt{2\pi\Omega^\alpha_\Lambda}}\,\exp\left(-\frac{\vp^2}{2\Omega^\alpha_\Lambda}\right)\,,
\ee
where
\be 
\Omega^\alpha_\Lambda = \left(\frac32-\frac1\alpha\right) \Omega_\Lambda\,,
\ee
and the tadpole corrections represented by the ellipses in \eqref{topFull} and \eqref{compGen} must now be computed with the propagators \eqref{hpalph} -- \eqref{hhalph}. All of the properties of the coefficient functions, reviewed in sec. \ref{sec:sollinear}, then go through unchanged. In particular the coefficient functions have an amplitude suppression scale $\Lambda_\p$ which can be chosen common to all of them, independent of the monomial, and such that they trivialise in the large amplitude suppression scale limit (\ref{flat},\ref{flatp}). The BRST representatives \eqref{Gsplit} -- \eqref{Gonezero} are the same as before since these are computed without gauge fixing (in particular this means they can still be taken to depend only on the minimal set). The same applies to the analysis in secs. \ref{sec:solCME} and \ref{sec:BRSTexact}. The only change is to the one-loop tadpole correction \eqref{Goneq}. The second order analysis of the conformal sector \cite{secondconf} also goes through unchanged since this effectively relies on dimensional analysis and general properties of the coefficient functions. This means that once again the renormalized trajectory can smoothly enter the diffeomorphism subspace as discussed in sec. \ref{sec:evaluation}. Once inside we have returned to the standard framework where the gauge parameter dependence is absent on shell and well understood off shell, as we already discussed above.

We have thus shown that the structure we have developed changes only in inessential ways for any value of the gauge fixing parameter $\alpha$ in the range \eqref{validalpha}.
It is interesting to examine what happens in the gap excluded by the inequalities \eqref{validalpha}. If $\alpha=-2$ (or $\alpha=\frac23$), the flow equation \eqref{flowalpha} no longer has a dependence on $h_{\mu\nu}$ (or $\ph$). The eigenoperator equation in this case is no longer of Sturm-Liouville type and some different notion of convergence over eigenoperators would have to be formulated \cite{\morri,\morrii}. If $-2<\alpha<0$ the sign on the right hand side of \eqref{flowalpha} is negative for both $\ph$ and $h_{\mu\nu}$. In this case there is no longer convergence when expanding over polynomials in $h_{\mu\nu}$ \cite{\morri,Dietz:2016gzg}. A sensible Wilsonian flow would require generalising the  $\dd{\Lambda}{n}$ to $h_{\mu\nu}$ fields. This opens up new possibilities for quantisation schemes which go beyond the present investigation. The choice $\alpha=0$ is Landau gauge and is singular; to define it requires taking a limit from $\alpha\ne0$. Finally if $0<\alpha<2/3$, both signs are positive on the right hand side of \eqref{flowalpha}. This does not mean we have removed the instability. One can verify that the tadpole integral $\int_p \langle R^{(1)}(p)\, R^{(1)}(-p)\rangle$ is negative and independent of $\alpha$. The instability remains, but in higher derivative terms. Like unimodular gravity, \cf sec. \ref{sec:discussion} and \cite{\morrii}, the consequences of instability for Wilsonian RG are less straightforward to analyse and go beyond the present investigation.
}

%\section{h sector}
%\label{sec:hsector}

\section{Summary and Conclusions}
\label{sec:conclusions}

In Euclidean signature the Einstein-Hilbert action is unbounded from below. This so-called conformal factor instability \cite{Gibbons:1978ac} means that the partition function for quantum gravity makes no sense without further modification. The authors of ref. \cite{Gibbons:1978ac}  proposed to solve this by analytically continuing the conformal factor along the imaginary axis. However the Wilsonian exact RG flow equation still makes sense in the presence of this instability \cite{Reuter:1996,\morri} and anyway provides a more powerful route to define the continuum limit. Nevertheless the instability has a profound effect on RG properties. We find that  flows close to the Gaussian fixed point, involving otherwise arbitrary functions of the conformal factor amplitude, $\vp$,  remain well defined only if expanded over a novel tower of increasingly relevant operators $\dd\Lambda{n}$ ($n=0,1,\cdots$) \cite{\morri}. Everything in the new quantisation just follows from this observation. 

The result is the renormalized trajectory sketched in fig. \ref{fig:flow}. Although at first sight this looks like the standard picture for a perturbative continuum limit, an important difference is  that the upper part lies outside the diffeomorphism invariant subspace where the corresponding BRST invariance (or rather modified Slavnov-Taylor identities) can be respected.  The quantisation is thus defined ``off space-time'' \cite{first} in the upper part of the renormalized trajectory. In this part, the interactions involve traceless fluctuations $h_{\mu\nu}$ and the conformal factor $\vp$, acting as separate fields.
The dynamical metric $g_{\mu\nu}$, which combines these as stipulated by diffeomorphism invariance, only comes together inside the diffeomorphism invariant subspace and does not make sense as a concept outside this subspace.

In ref. \cite{secondconf} we solved for the renormalized trajectory for pure quantum gravity at second order in perturbation theory and showed that, for the underlying coupling constants in appropriate domains, the trivialisation conditions can be satisfied. In this paper we have shown that it is then indeed possible for the renormalized trajectory  to enter the diffeomorphism invariant subspace. We then solved for its subsequent evolution, in particular for the limit $\Lambda\cu\to0$ where we recover the physical amplitudes. As we saw, the result is equivalent to solving for pure quantum gravity at one loop and $O(\kappa^2)$ in standard perturbation theory. It is not so surprising therefore that we also find that effective parameters are left behind associated to logarithmically running terms at this order, and that for pure quantum gravity these are not physical because they can be absorbed by reparametrisations. 

Beyond $O(\kappa^2)$ in pure quantum gravity and/or after including matter or a cosmological constant, it is no longer true in the usual treatment that logarithmic divergences can be absorbed by reparametrisation. Instead they force the introduction of new couplings order by order in the loop expansion. The main question then is whether in this new quantisation one similarly finds that ultimately an infinite number of diffeomorphism invariant effective couplings are required, introduced order by order in perturbation theory. If this is the case, it appears one is left with a genuine entirely consistent continuum theory of perturbative quantum gravity which, unfortunately for its phenomenology, is controlled by an infinite number of couplings.

Actually the precise correspondence, of pure quantum gravity at second order in the new quantisation, to standard quantisation of effective quantum gravity at one-loop and $O(\kappa^2)$,  is somewhat of an accident, see below \eqref{partLASS}. The interactions in the upper part of the renormalized trajectory are second order in couplings, but non-perturbatively quantum, and thus involve a sum to all loops over tadpoles and melonic Feynman diagrams. On entering the diffeomorphism invariant subspace, this collapses to something that can be reinterpreted as finite order in $\hbar$. Furthermore at second order, the order in $\hbar$ amounts to one loop in the loop expansion. At higher orders it looks like the large-$\Lambda_\p$ limit may differ from the standard solution in that not all contributions perturbative in $\hbar$ are reproduced up to the maximum number of loops that appear. It seems therefore that higher order will imply a finite reordering of the loop-wise expansion, but it is not clear that this has a physical consequence. From third order onwards, the first-order underlying couplings (that parametrise the first order vertices) will run \cite{secondconf}. This may lead to restrictions on matching into the diffeomorphism invariant subspace. On the other hand, since there is no corresponding running of $\kappa$ in standard quantisation, we expect the running to effectively freeze out on entering the diffeomorphism invariant subspace, as a consequence of the trivialisation conditions.

Finally in sec. \ref{sec:missing} we noted that the particular parabolic properties of these flow equations mean that solutions are typically singular when evolved in either the IR or UV directions, once one works with a space of solutions that is non-polynomial in both the quantum fields $h_{\mu\nu}$ and $\vp$. Non-perturbatively in $\kappa$ the solutions must indeed be non-polynomial in these quantum variables, as forced by diffeomorphism invariance via the mST. We uncovered hints that this property provides a non-perturbative mechanism which fixes the free parameters down to just $\kappa$ and the cosmological constant. It would appear to be sufficient to have this mechanism at work entirely within the diffeomorphism invariant subspace. Then the theory can be defined after all by working solely within this space. But then the understanding of how the continuum limit is achieved would be very different from the Wilsonian one, since it would not be in terms of a renormalized trajectory emanating from an ultraviolet fixed point.

%\newpage
%\mbox{}
%\newpage

\section*{Acknowledgments}

AM and MPK acknowledge support via an STFC PhD studentship. TRM acknowledges support from STFC through Consolidated Grant ST/P000711/1.

%\vfill
%\newpage 

\appendix

\section{Computing Taylor expanded IR regulated momentum integrals}
\label{app:derivexp}

To compute derivative expansions such as those that appear for level-one \eqref{Bderiv}, level-zero  \eqref{Aderiv} and in the mST correction term \eqref{Fderiv}, we Taylor expand their integrands in the external momentum $p_\mu$. We use the $d$-dimensional equivalent of the integrands (\ref{Bint},\ref{Aint},\ref{Fint}) displayed in the paper, constructed from using the $d$-dimensional propagators \eqref{HH}--\eqref{cc} attached to the $d$-dimensional $\cG_1$ described at the end of sec. \ref{sec:sollinear}. To be concrete we describe how to treat $\mathcal{B}_{\mu\nu\alpha}(p,\Lambda)$ and $\mathcal{A}_{\mu\nu\alpha\beta}(p,\Lambda)$ in the following. We comment on the slight differences for $\mathcal{F}_{\mu\nu\alpha}(p,\Lambda)$ later.
The Taylor expansion coefficients involve the integrals
\be 
\label{TaylorExpansionCoeffs}
\int_q \frac{q_{\mu_1}q_{\mu_2}\cdots q_{\mu_{2n}}}{q^{2r}}\, \bar{C}(q^2/\Lambda^2)\, \bar{C}^{(m)}(q^2/\Lambda^2)\,,
\ee
for some non-negative integers $m,n,r$, with normalisation of the measure as in \eqref{defs}.
Here $\bar{C}(u) = 1\cu-C(u)$ is the IR cutoff function, and $\bar{C}^{(m)}(u)$ is its $m^\text{th}$ differential, where  $u\cu=q^2/\Lambda^2$. Now $d$-dimensional rotational invariance ensures that the integral vanishes unless the numerator has even powers of $q$ and moreover it allows us to reduce the latter to a scalar integral using 
\be 
\label{iterativeaverage}
q_{\mu_1}q_{\mu_2}\cdots q_{\mu_{2n}} \equiv q^{2n} \prod_{k=1}^n \frac1{d+2(k-1)} \sum_{\text{pairs}} \delta_{\mu_{\sigma_1}\mu_{\sigma_2}}\delta_{\mu_{\sigma_3}\mu_{\sigma_4}}\cdots 
\delta_{\mu_{\sigma_{2n-1}}\mu_{\sigma_{2n}}}\,,
\ee
where this formula is valid under the integral, and may be proved by iteration. The sum is over all ways of dividing the $2n$ indices into Kronecker-delta pairs. For these one-loop integrals in $d\cu=4\cu-2\epsilon$ dimensions, the worst we can get is a $1/\epsilon$ pole, therefore up to terms vanishing as $\epsilon\cu\to0$, \com{998.7} 
\be 
\int_q = \left(1+ \left[ 1-\gamma_E +\ln(4\pi/\Lambda^2) \right]\! \epsilon\,\right) 
\int^\infty_0\!\!\!\!\!\!du\, u^{1-\epsilon}\,.
\ee
The integrals are now reduced iteratively using integration by parts on those containing the highest differential $\bar{C}^{(m)}$. Following the philosophy of dimensional regularisation we choose $\epsilon\cu>0$ large enough such that we can always discard the UV limit (\aka surface term). The IR limit can also be discarded using the same philosophy, choosing $\epsilon\cu<0$ negative enough.\footnote{At high orders in the derivative expansion this allows us to discard the lower boundary, $\lim_{\epsilon\to0} u^{-k-\eps}\, \bar{C}^{(m)}\bar{C}^{(n)}$ for any positive integers $k,m,n$. This could also be assured by choosing $C$ such that it has vanishing Taylor expansion to all orders at $u\cu=0$ (known as a  ``bump'' function). In practice in the cases dealt with in sec. \ref{sec:standard} the lower limit can be discarded anyway thanks to the presence of $\bar{C}(u)$ and/or positive integer powers of $u$.} After this we analytically continue $\epsilon$ to the neighbourhood of $\epsilon\cu=0$ in the usual way. As a simple but instructive example we thus have the identity
\be 
\label{example}
\int^\infty_0\!\!\!\!\!\!du\, u^{-\epsilon}\, \bar{C} \frac{d}{du}\bar{C} = \frac{\epsilon}{2}\int^\infty_0\!\!\!\!\!\!du\, u^{-1-\epsilon}\, \bar{C}^2\,.
\ee
At the end of the process, provided at least one of the $\bar{C}$ is differentiated, the integral is in fact both UV and IR regulated by the cutoff function, and thus $\epsilon\cu\to0$ can be safely taken. The integrals that require more care are those that are only IR regulated which thus take the form 
\beal 
\int^\infty_0\!\!\!\!\!\!du\, u^{n-\epsilon}\, \bar{C}^2(u) &= \int^1_0\!\!\!\!du\, u^{n-\epsilon}\, \bar{C}^2\ + 
\int^\infty_1\!\!\!\!\!\!du\, u^{n-\epsilon}\, \left(\bar{C}^2-1\right)\  + \int^\infty_1\!\!\!\!\!\!du\, u^{n-\epsilon}\,,\nn\\
&= \int^1_0\!\!\!\!du\, u^{n}\, \bar{C}^2\ + 
\int^\infty_1\!\!\!\!\!\!du\, u^{n}\, C(C-2)\  -\frac1{n+1-\epsilon}+O(\epsilon)\,,
\label{care}
\eeal
for some integer $n$. Splitting the integral into three parts as in the first line, we see that the first two parts are both IR and UV regulated for any $n$ and thus $\epsilon\cu\to0$ can be safely taken. The final integral gives the last term on discarding the upper limit. 

As a simple example consider the case $n\cu=-1$. This appears on the RHS of \eqref{example}. Substituting \eqref{care} and taking the limit $\epsilon\cu\to0$ one finds the answer  $\half$. In this case it is straightforward to derive this directly from \eqref{example} at $\epsilon\cu=0$, since the LHS is then a total derivative and the answer $\half$ is recovered from the UV boundary.  However applying dimensional regularisation to all cases including the more involved (\ref{TaylorExpansionCoeffs},\ref{iterativeaverage}) cases, ensures that results are not subject to momentum routing (equivalently surface term) ambiguities.

In \eqref{care}, apart from the case $n\cu=-1$  which, if it has non-vanishing coefficient, is subtracted using $\overline{\text{MS}}$ \cf comments above \eqref{flowtwocl}, the $\epsilon\cu\to0$ limit of the last term can also now be safely taken. It then just cancels the cutoff-independent contribution in the first integral on the RHS, thus
\be 
\int^\infty_0\!\!\!\!\!\!du\, u^{n-\epsilon}\, \bar{C}^2(u) = \int^\infty_0\!\!\!\!\!\!du\, u^{n}\, C(C-2) \ +O(\epsilon)\,,\qquad (n\ne-1)\,,
\ee
which we could have derived directly from substituting $\bar{C}\cu=1\cu-C$, and discarding the cutoff independent piece as would be done as standard in dimensional regularisation (despite the fact that the integral is strictly speaking ill-defined for any $\epsilon$).

%\vfill
%\newpage 

\bibliographystyle{hunsrt}
\bibliography{references} %%from now on (14/7/15) this is the global references list!

\begin{thebibliography}{10}

\bibitem{tHooft:1974toh}
Gerard 't~Hooft and M.~J.~G. Veltman.
\newblock {One loop divergencies in the theory of gravitation}.
\newblock {\em Ann. Inst. H. Poincare Phys. Theor.}, A20:69--94, 1974.

\bibitem{Goroff:1985sz}
Marc~H. Goroff and Augusto Sagnotti.
\newblock {Quantum Gravity at Two Loops}.
\newblock {\em Phys. Lett.}, B160:81--86, 1985.

\bibitem{Goroff:1985th}
Marc~H. Goroff and Augusto Sagnotti.
\newblock {The Ultraviolet Behavior of Einstein Gravity}.
\newblock {\em Nucl. Phys.}, B266:709--736, 1986.

\bibitem{vandeVen:1991gw}
Anton E.~M. van~de Ven.
\newblock {Two loop quantum gravity}.
\newblock {\em Nucl. Phys.}, B378:309--366, 1992.

\bibitem{Wilson:1973}
K.G. Wilson and John~B. Kogut.
\newblock {The Renormalization group and the epsilon expansion}.
\newblock {\em Phys.Rept.}, 12:75--200, 1974.

\bibitem{Wegner:1972my}
Franz~J. Wegner.
\newblock {Corrections to scaling laws}.
\newblock {\em Phys. Rev.}, B5:4529--4536, 1972.

\bibitem{Morris:2018mhd}
Tim~R. Morris.
\newblock {Renormalization group properties in the conformal sector: towards
  perturbatively renormalizable quantum gravity}.
\newblock {\em JHEP}, 08:024, 2018, 1802.04281.

\bibitem{Gibbons:1978ac}
G.W. Gibbons, S.W. Hawking, and M.J. Perry.
\newblock {Path Integrals and the Indefiniteness of the Gravitational Action}.
\newblock {\em Nucl.Phys.}, B138:141, 1978.

\bibitem{Kellett:2018loq}
Matthew~P. Kellett and Tim~R. Morris.
\newblock {Renormalization group properties of the conformal mode of a torus}.
\newblock {\em Class. Quant. Grav.}, 35(17):175002, 2018, 1803.00859.

\bibitem{Morris:2018upm}
Tim~R. Morris.
\newblock {Perturbatively renormalizable quantum gravity}.
\newblock {\em Int. J. Mod. Phys.}, D27(14):1847003, 2018, 1804.03834.

\bibitem{Morris:2018axr}
Tim~R. Morris.
\newblock {Quantum gravity, renormalizability and diffeomorphism invariance}.
\newblock {\em SciPost Phys.}, 5:040, 2018, 1806.02206.

\bibitem{first}
Alex Mitchell and Tim~R. Morris.
\newblock {The continuum limit of quantum gravity at first order in
  perturbation theory}.
\newblock {\em JHEP}, 06:138, 2020, 2004.06475.

\bibitem{secondconf}
Tim~R. Morris.
\newblock {The continuum limit of the conformal sector at second order in
  perturbation theory}.
\newblock 2020, 2006.05185.

\bibitem{Ellwanger:1994iz}
Ulrich Ellwanger.
\newblock {Flow equations and BRS invariance for Yang-Mills theories}.
\newblock {\em Phys. Lett.}, B335:364--370, 1994, hep-th/9402077.

\bibitem{Kadanoff:1966wm}
L.~P. Kadanoff.
\newblock {Scaling laws for Ising models near T(c)}.
\newblock {\em Physics}, 2:263--272, 1966.

\bibitem{Barnich:1993vg}
Glenn Barnich and Marc Henneaux.
\newblock {Consistent couplings between fields with a gauge freedom and
  deformations of the master equation}.
\newblock {\em Phys. Lett.}, B311:123--129, 1993, hep-th/9304057.

\bibitem{Boulanger:2000rq}
Nicolas Boulanger, Thibault Damour, Leonardo Gualtieri, and Marc Henneaux.
\newblock {Inconsistency of interacting, multigraviton theories}.
\newblock {\em Nucl. Phys.}, B597:127--171, 2001, hep-th/0007220.

\bibitem{ZinnJustin:2002ru}
Jean Zinn-Justin.
\newblock {Quantum field theory and critical phenomena}.
\newblock {\em Int. Ser. Monogr. Phys.}, 113:1--1054, 2002.

\bibitem{Igarashi:2019gkm}
Yuji Igarashi, Katsumi Itoh, and Tim~R. Morris.
\newblock {BRST in the Exact RG}.
\newblock {\em PTEP}, 2019(10):103B01, 2019, 1904.08231.

\bibitem{Batalin:1981jr}
I.~A. Batalin and G.~A. Vilkovisky.
\newblock {Gauge Algebra and Quantization}.
\newblock {\em Phys. Lett.}, 102B:27--31, 1981.
\newblock [,463(1981)].

\bibitem{Batalin:1984jr}
I.~A. Batalin and G.~A. Vilkovisky.
\newblock {Quantization of Gauge Theories with Linearly Dependent Generators}.
\newblock {\em Phys. Rev.}, D28:2567--2582, 1983.
\newblock [Erratum: Phys. Rev.D30,508(1984)].

\bibitem{Nicoll1977}
J.~F. Nicoll and T.~S. Chang.
\newblock {An Exact One Particle Irreducible Renormalization Group Generator
  for Critical Phenomena}.
\newblock {\em Phys. Lett.}, A62:287--289, 1977.

\bibitem{Wetterich:1992}
Christof Wetterich.
\newblock {Exact evolution equation for the effective potential}.
\newblock {\em Phys.Lett.}, B301:90--94, 1993.

\bibitem{Morris:1993}
Tim~R. Morris.
\newblock {The Exact renormalization group and approximate solutions}.
\newblock {\em Int.J.Mod.Phys.}, A 09:2411--2450, 1994, hep-ph/9308265.

\bibitem{Weinberg:1976xy}
Steven Weinberg.
\newblock {Critical Phenomena for Field Theorists}.
\newblock In {\em {14th International School of Subnuclear Physics:
  Understanding the Fundamental Constitutents of Matter Erice, Italy, July
  23-August 8, 1976}}, page~1, 1976.

\bibitem{Morris:2015oca}
Tim~R. Morris and Zo{\"e}~H. Slade.
\newblock {Solutions to the reconstruction problem in asymptotic safety}.
\newblock {\em JHEP}, 11:094, 2015, 1507.08657.

\bibitem{Bonini:1992vh}
M.~Bonini, M.~D'Attanasio, and G.~Marchesini.
\newblock {Perturbative renormalization and infrared finiteness in the Wilson
  renormalization group: The Massless scalar case}.
\newblock {\em Nucl. Phys.}, B409:441--464, 1993, hep-th/9301114.

\bibitem{Ellwanger1994a}
Ulrich Ellwanger.
\newblock {Flow equations for N point functions and bound states}.
\newblock {\em Z. Phys.}, C62:503--510, 1994, hep-ph/9308260.
\newblock [,206(1993)].

\bibitem{Morgan1991}
D.~Morgan.
\newblock {\em Quartet: Baryogenesis, Bubbles of False Vacuum, Quantum Black
  Holes, and the Renormalization Group}.
\newblock PhD thesis, University of Texas, Austin, 1991.

\bibitem{Gomis:1994he}
Joaquim Gomis, Jordi Paris, and Stuart Samuel.
\newblock {Antibracket, antifields and gauge theory quantization}.
\newblock {\em Phys. Rept.}, 259:1--145, 1995, hep-th/9412228.

\bibitem{Morris:2016nda}
Tim~R. Morris and Anthony W.~H. Preston.
\newblock {Manifestly diffeomorphism invariant classical Exact Renormalization
  Group}.
\newblock {\em JHEP}, 06:012, 2016, 1602.08993.

\bibitem{Stelle:1976gc}
K.~S. Stelle.
\newblock {Renormalization of Higher Derivative Quantum Gravity}.
\newblock {\em Phys. Rev.}, D16:953--969, 1977.

\bibitem{Morris:1998}
Tim~R. Morris.
\newblock {Elements of the continuous renormalization group}.
\newblock {\em Prog.Theor.Phys.Suppl.}, 131:395--414, 1998, hep-th/9802039.

\bibitem{WR}
F.~J. Wegner.
\newblock Some invariance properties of the renormalization group.
\newblock {\em J. Phys.}, C7:2098, 1974.

\bibitem{Dietz:2013sba}
Juergen~A. Dietz and Tim~R. Morris.
\newblock {Redundant operators in the exact renormalisation group and in the
  f(R) approximation to asymptotic safety}.
\newblock {\em JHEP}, 07:064, 2013, 1306.1223.

\bibitem{Percacci:2017fsy}
R.~Percacci.
\newblock {Unimodular quantum gravity and the cosmological constant}.
\newblock {\em Found. Phys.}, 48(10):1364--1379, 2018, 1712.09903.

\bibitem{Weinberg:1980}
S.~Weinberg.
\newblock {Ultraviolet Divergences In Quantum Theories Of Gravitation}.
\newblock {\em In Hawking, S.W., Israel, W.: General Relativity; Cambridge
  University Press}, pages 790--831, 1980.

\bibitem{Reuter:1996}
M.~Reuter.
\newblock {Nonperturbative evolution equation for quantum gravity}.
\newblock {\em Phys.Rev.}, D57:971--985, 1998, hep-th/9605030.

\bibitem{Bonanno:2020bil}
Alfio Bonanno, Astrid Eichhorn, Holger Gies, Jan~M. Pawlowski, Roberto
  Percacci, Martin Reuter, Frank Saueressig, and Gian~Paolo Vacca.
\newblock {Critical reflections on asymptotically safe gravity}.
\newblock 2020, 2004.06810.

\bibitem{Dietz:2016gzg}
Juergen~A. Dietz, Tim~R. Morris, and Zoe~H. Slade.
\newblock {Fixed point structure of the conformal factor field in quantum
  gravity}.
\newblock {\em Phys. Rev.}, D94(12):124014, 2016, 1605.07636.

\bibitem{Ambjorn:2015qja}
Jan Ambjrn, Daniel~N. Coumbe, Jakub Gizbert-Studnicki, and Jerzy Jurkiewicz.
\newblock {Signature Change of the Metric in CDT Quantum Gravity?}
\newblock {\em JHEP}, 08:033, 2015, 1503.08580.
\newblock [JHEP08,033(2015)].

\bibitem{Chaney:2015mfa}
A.~Chaney, Lei Lu, and A.~Stern.
\newblock {Lorentzian Fuzzy Spheres}.
\newblock {\em Phys. Rev.}, D92(6):064021, 2015, 1506.03505.

\bibitem{Steinacker:2017vqw}
Harold~C. Steinacker.
\newblock {Cosmological space-times with resolved Big Bang in Yang-Mills matrix
  models}.
\newblock {\em JHEP}, 02:033, 2018, 1709.10480.

\bibitem{Stern:2018wud}
A.~Stern and Chuang Xu.
\newblock {Signature change in matrix model solutions}.
\newblock {\em Phys. Rev.}, D98(8):086015, 2018, 1808.07963.

\bibitem{Perry:1993ry}
Malcolm~J. Perry and Edward Teo.
\newblock {Nonsingularity of the exact two-dimensional string black hole}.
\newblock {\em Phys. Rev. Lett.}, 70:2669--2672, 1993, hep-th/9302037.

\bibitem{Bojowald:2016itl}
Martin Bojowald and Suddhasattwa Brahma.
\newblock {Signature change in two-dimensional black-hole models of loop
  quantum gravity}.
\newblock {\em Phys. Rev.}, D98(2):026012, 2018, 1610.08850.

\bibitem{Bojowald:2018xxu}
Martin Bojowald, Suddhasattwa Brahma, and Dong-han Yeom.
\newblock {Effective line elements and black-hole models in canonical loop
  quantum gravity}.
\newblock {\em Phys. Rev.}, D98(4):046015, 2018, 1803.01119.

\bibitem{ZinnJustin:1974mc}
Jean Zinn-Justin.
\newblock {Renormalization of Gauge Theories}.
\newblock {\em Lect. Notes Phys.}, 37:1--39, 1975.

\bibitem{ZinnJustin:1975wb}
Jean Zinn-Justin.
\newblock {Renormalization Problems in Gauge Theories}.
\newblock In {\em {Functional and Probabilistic Methods in Quantum Field
  Theory. 1. Proceedings, 12th Winter School of Theoretical Physics, Karpacz,
  Feb 17-March 2, 1975}}, pages 433--453, 1975.

\bibitem{Kawai:1993mb}
Hikaru Kawai, Yoshihisa Kitazawa, and Masao Ninomiya.
\newblock {Ultraviolet stable fixed point and scaling relations in
  (2+epsilon)-dimensional quantum gravity}.
\newblock {\em Nucl. Phys.}, B404:684--716, 1993, hep-th/9303123.

\bibitem{Eichhorn:2013xr}
Astrid Eichhorn.
\newblock {On unimodular quantum gravity}.
\newblock {\em Class. Quant. Grav.}, 30:115016, 2013, 1301.0879.

\bibitem{Nink:2014yya}
Andreas Nink.
\newblock {Field Parametrization Dependence in Asymptotically Safe Quantum
  Gravity}.
\newblock {\em Phys. Rev.}, D91(4):044030, 2015, 1410.7816.

\bibitem{Percacci:2015wwa}
Roberto Percacci and Gian~Paolo Vacca.
\newblock {Search of scaling solutions in scalar-tensor gravity}.
\newblock {\em Eur. Phys. J.}, C75(5):188, 2015, 1501.00888.

\bibitem{Percacci:2016arh}
Roberto Percacci and Gian~Paolo Vacca.
\newblock {The background scale Ward identity in quantum gravity}.
\newblock {\em Eur. Phys. J.}, C77(1):52, 2017, 1611.07005.

\bibitem{Rosten:2010pc}
Oliver~J. Rosten.
\newblock {Equivalent Fixed-Points in the Effective Average Action Formalism}.
\newblock {\em J. Phys.}, A44:195401, 2011, 1010.1530.

\end{thebibliography}

\end{document}